# Single Shot Solar Spectroscopic Design Aspects

*A thesis submitted to the*

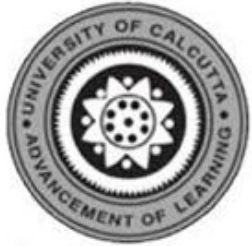

**Department of Applied Optics & Photonics**
**University of Calcutta**

In the partial fulfillment of the requirements
for the award of the degree of
**Master of Technology**
**In**
**Astronomical Instrumentation**

By

**Harsh Mathur**

Under the guidance of

**Dr. K. Nagaraju**
**Dr. K. Sankarasubramanian**
**Dr. Shibu K. Mathew**

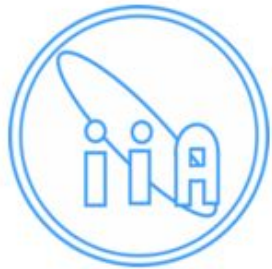

**Indian Institute of Astrophysics, Bangalore**

# Declaration

I hereby declare that the thesis entitled "Single Shot Solar Spectroscopic Design Aspects" submitted to the University of Calcutta in the Department of Applied Optics and Photonics, in partial fulfillment of the requirements for the award of the degree of Master of Technology, is a result of the project work carried out by me at Indian Institute of Astrophysics, Bangalore, under the supervision of Dr K. Nagaraju (IIA), Dr K. Sankarsubramanian (ISRO) and Dr. Shibu K. Mathew (USO). The results presented herein have not been subject to scrutiny, by any university or institute, for the award of a degree, diploma, associateship or fellowship whatsoever.

<div style="text-align: right;">
______________________________

Harsh Mathur
</div>

Date:
Indian Institute of Astrophysics Bangalore-560034
India

# Certificate

This is to certify that the thesis entitled "Single Shot Solar Spectroscopic Design Aspects" submitted to the University of Calcutta by Mr. Harsh Mathur in partial fulfillment of the requirements for the award of the degree of Master of Technology in Astronomical Instrumentation is based on the results of the project work carried out by him under my supervision and guidance, at the Indian Institute of Astrophysics. This thesis has not been submitted for the award of any degree, diploma, associateship, fellowship, etc. of any university or institute.

<div style="text-align:right">

\_\_\_\_\_\_\_\_\_\_\_\_\_\_\_\_\_\_\_\_\_\_\_\_\_\_\_\_

Dr K. Nagaraju

\_\_\_\_\_\_\_\_\_\_\_\_\_\_\_\_\_\_\_\_\_\_\_\_\_\_\_\_

Dr K. Sankarsubramanian

</div>

Date:
Indian Institute of Astrophysics Bangalore-560034
India

> "There's no such thing as perfect writing,
> just like there's no such thing as perfect despair."
> --Haruki Murakami

# Acknowledgments

I'd like to express my sincere gratitude to Dr. K. Nagaraju, Dr. K. Sankarsubramanian, Dr. Shibu K. Mathew for taking me up with this project. I'd like to acknowledge and thank Nagaraju for keeping up with my impatient nature and helping me work on being more patient. I also like to acknowledge and express my gratitude for taking the time to clear my doubts regarding the project and concepts in radiative transfer and spectropolarimetry. I'd like to thank Sankarsubramanian Sir for the detailed discussions about the DKIST snapshot spectrograph design and other discussions on relevant topics. I'd like to thank Shibu for inviting me to Udaipur and helping me with instrumentation, camera programs, serial port interfacing, etc. I'd also like to thank Dr. Raja Bayanna for helping me optimize Zeemax designs and designing achromats and helping me with setting up the instrument. I'd like to thank Dr. Dipankar Banerjee for the discussions about solar physics topics including but not limited to the solar atmosphere, waves, etc. I'd like to thank Sriram Sir for his invaluable insight in modeling and analyzing the performance of microlens arrays. I'd also like to acknowledge and thank Totan Sir for his constant help and involvement in designing the achromats.

Next, I'd like to thank my seniors Dr. Hemanth Pruthvi for helping me with spectro-polarimetry concepts and software interfacing with instruments. I'd like to thank my senior Dr. Mohan for the explanations about the snapshot spectrograph data reduction process and clearing other doubts as I had. I'd also like to thank my senior Souvik Bose for the discussions and explanations about sunspots and radiative transfer and for motivating me. I'd like to thank my senior Phanindra for helping me in the Optics lab with the instrument. I'd like to thank my senior Dr. Subhamoy for his ideas in optimizing the Zeemax design and image processing. I'd like to thank my senior Ritesh Patel for helping me time to time with reports and presentations. I'd like to thank Bibhuti and Kamlesh for the helpful discussions and explanations. I'd like to thank my batchmates Sarthak, Kshitij, Soumya, Pruthvi, Bharat and Vishnu for their constant involvement and support. I'd also like to thank my friends Ankit and Siddharth for moral support and being the pillars of strength while I made this career move.

I'm grateful towards Dr. Ravindra Banyal, Dr. Mousami Das and Dr. Subinoy Das for their invaluable classes. I'm grateful towards our BGS coordinator Dr. Ravindra and BGS Chair Dr. Aruna Goswami and BGS assistant Sankar Sir for their support.

I'd like to thank the computer support team to help me with troubleshooting systems and software whenever needed.

I'd also like to thank faculties from Calcutta University namely Prof. K. Bhattachaya, Prof. A. Ghosh, Prof. L.N. Hazra, Prof. A.B.Ray, Prof. S. Sarkar, Prof. A. Chakrabarty, Dr. R. Chakraborty, Dr. M. Ray for their wonderful teaching. I'd like to thank Dr. N. Chakraborty for their classes.

Above all, I'd like to thank my parents for their constant support, love, and blessings. I'd like to thank my brother Vibhor for being a pillar of strength and taking the responsibility of my parents for me to freely pursue the studies.

# Abstract

High time-cadence Spectro-Polarimetry allows the feasibility of studying magnetic field evolution coupled with the plasma flows. Such a high cadence solar spectro-polarimetry if developed will allow one to study magnetic field evolution in eruptive processes like solar flares, prominence eruptions, etc. A single shot solar spectroscopy was recently demonstrated at Multi-application Solar Telescope (MAST) at Udaipur Solar Observatory. The snapshot spectroscopy is performed by sampling the pupil plane using the lenslet array to get multiple images of the field of view (FOV), which are then collimated and the collimated beam is made to pass through an FP Etalon in collimated mode. As the distance from the FP axis increases, the peak transmitted wavelength shift towards the bluer side. Using a pre-filter with a full width half maximum (FWHM) less than the free spectral range (FSR) of FP, combined with an imaging lens, we can get multiple images of FOV on image plane with a blue shift in spectra as we move radially outwards from the optical axis.

We have made an optimized Zeemax design utilizing the above concept for Multi-application Solar Telescope (MAST) at Udaipur Solar Observatory and also demonstrated the concept using available components and collected sample data.

We have made a Zeemax design of the Snapshot Spectrograph for the Daniel K. Inouye Solar Telescope (DKIST), carried out tolerance analysis through Monte-Carlo simulations and presented the Spot diagrams and PSF for the worst and best case scenarios.

We have also attempted to demonstrate snapshot spectro-polarimetric concepts in IIA Optics Lab. We have used Stokes definition polarimeter i.e. a linear analyzer and a quarter wave plate combination as a polarimeter. We have used He-Ne laser followed by a linear polarizer as a light source and collimated it from a 50-micron pinhole, the collimated beam is incident upon the lenslet array making multiple images of the pinhole, images are further collimated and made to pass through the FP and imaged at the detector. We have presented the results of this experiment i.e. Stokes parameters.

# Contents

# List of Figures

# List of Tables

# 1 Introduction

As we get better telescopes we get better resolution and time cadence, thus we are able to resolve and discover more magnetic features. Spectroscopy and spectropolarimetry are widely used techniques to understand the dynamics of the magnetic field and the evolution of different magnetic features both spatially and temporally. One example is magnetic bright points (BP) which exist in 200 km scales and have very fast evolving times. Magnetic bright points are small scale structures thought to trace intense kG magnetic concentrations. The BP motions can be used to measure the dynamic properties of magnetic flux tubes and their interaction with granular plasma. Random motions inside BPs can create Alfven wave turbulence, which dissipates the waves in a coronal loop.

Another interesting magnetic phenomenon is solar flares. Solar flares are a sudden release of energy in the solar atmosphere, which is seen as intense variations in brightness. The evidence suggests that solar flares are magnetic reconnection events when the stored or built up magnetic energy is thrown out, which is why most solar flares are seen in active regions. Solar flares often but not always are followed by CME eruptions. Solar flares are classified as A, B, C, M X in increasing order of peak X-ray flux. Solar flares happen on timescales of minutes to hours.

To understand the magnetic phenomenon and associated dynamics, we need spectropolarimetric information at very fast time scales over the full field of view (FOV). The traditional methods of spectroscopy although widely used have some limitations. As described in the following paragraphs, the technique has low time cadence, it can either capture FOV (2D spatial information) and scans over the line profile or captures the line information in one exposure but scans over the full field. The proposed snapshot spectroscopy technique neither scans in wavelength nor scans on the FOV but captures line information and spatial information in a single exposure.

## 1.1 Traditional Methods of Spectroscopy

### 1.1.1 Push broom scanning

A single spatial point is taken from a 2D spatial region and its spectrum is obtained through dispersers like prism or grating. Then a different spatial point is taken from the field to repeat the process. The scanning mechanism involves movement in x and y-direction to generate a spectral map of the whole field point by point which when combined forms a hyperspectral cube of the field. (Figure 1.1)

### 1.1.2 1D line scanning

This is a slit and grating-based spectrograph and the mechanism involves a slit which samples a 1D region of the 2D field and generates a wavelength map along with all the points in that 1D field. This slit is moved or raster scanned over the 2D field to generate a hypercube image of the field. (Figure 1.1)

### 1.1.3 Wavelength scanning

This method involves obtaining spatial information of a 2D image at a particular wavelength using narrow band filters or Fabry Perot interferometer (FP). The hyperspectral cube is obtained by scanning across the line by taking several images by tuning the filter or FP. (Figure 1.1)

## 1.2 Hyperspectral Image

Spectroscopy provides a plethora of information about the FOV. A spectral image of the field consists of two spatial dimensions (say x-direction and y-direction) and one spectral dimension (wavelength- λ). A data cube which is made up of two spatial and one spectral dimension is called a hyperspectral cube or Hyperspectral Image (HSI) and the dimensions are represented as [x, y, λ] as shown in Figure 1.1.

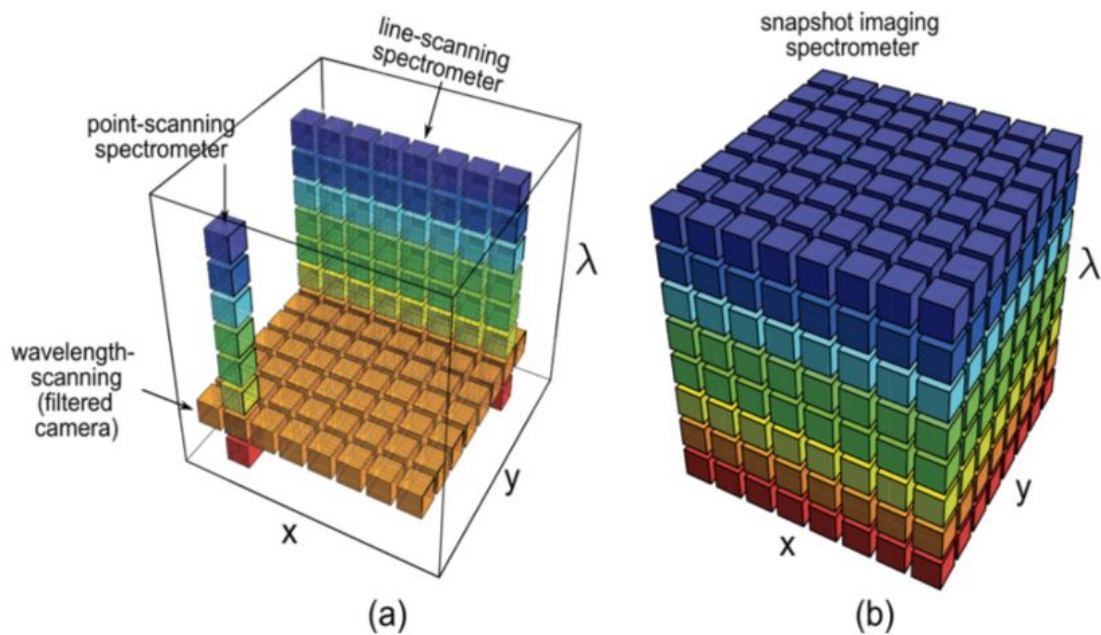

Fig 1.1. A hyperspectral data cube with the three-axis representing two spatial dimensions and one spectral dimension. (a) Represents the three scanning mechanisms commonly used to generate a hyperspectral cube. (b) Represents a hypercube of an object consisting of the three dimensions [x, y, λ] obtained in a snapshot

It is always necessary to generate the spectral information of an object before its spectral characteristics change or the object moves out of the field of view.

## 1.3 Snapshot Hyperspectral Techniques in Astronomy

Snapshot spectroscopy is a method of obtaining the hyperspectral cube in a single image frame i.e. a 3D data cube of [x, y, λ] is generated using a 2D detector plane in a single exposure.

### 1.3.1 Snapshot Spectroscopy with faceted mirrors

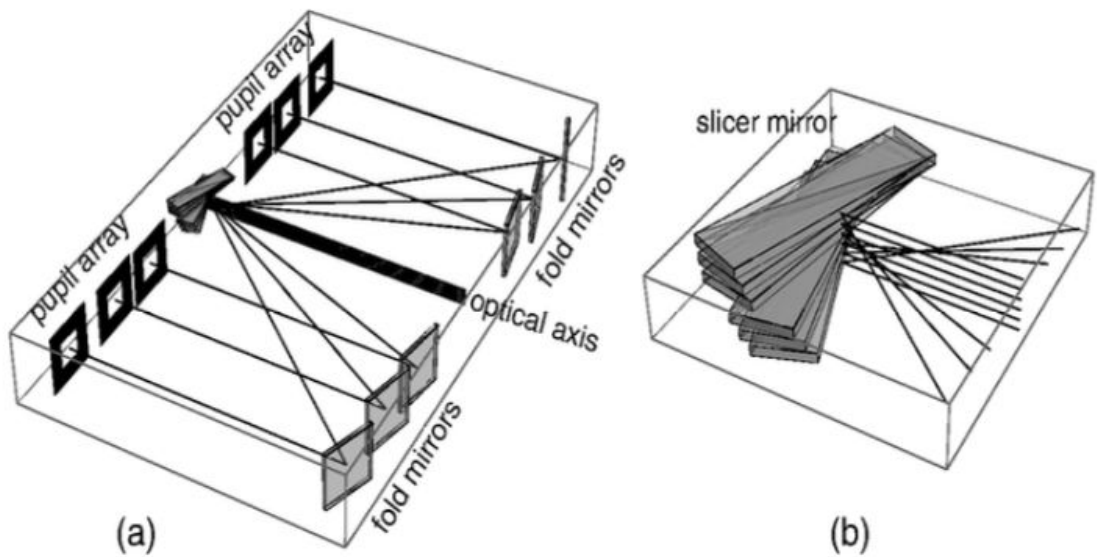

Fig 1.2 Facet mirror slicers divide the image and divert it into multiple spectrographs.

As shown in Figure 1.2, a set of closely spaced mirrors with slightly different angles such that an object whose image formed on this mirror is divided into several small regions. Each region is fed to separate spectrograph units and a spectrum is obtained. Then the divided regions and their respective spectra are combined to form Hyperspectral Image of the object.

## 1.3.2 Snapshot Spectroscopy with lenslets on the image plane

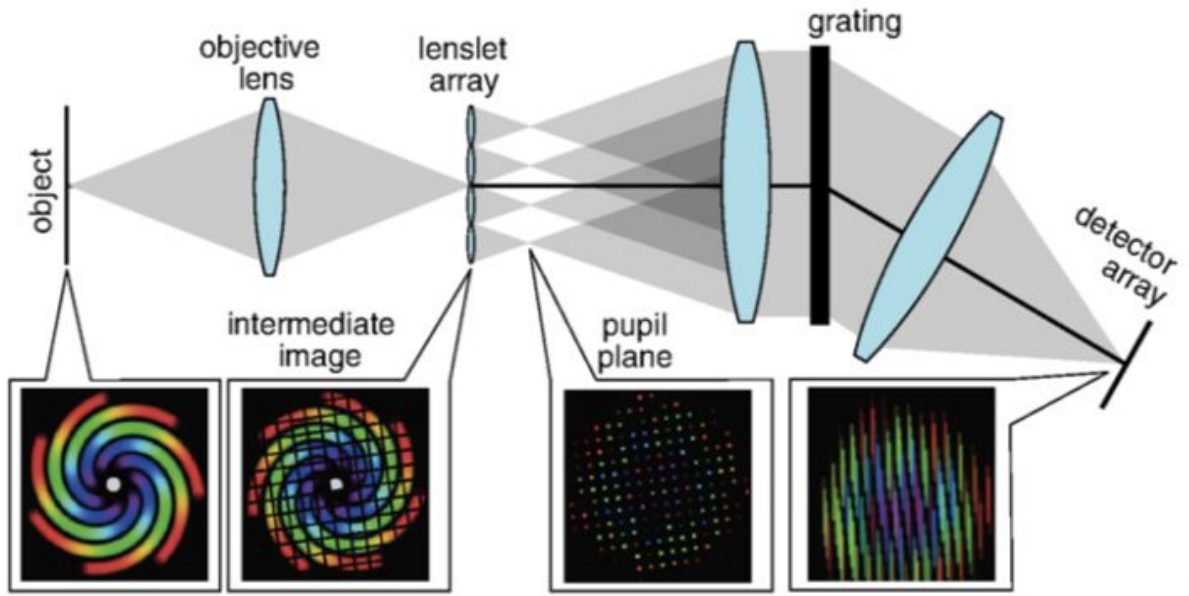

Fig 1.3 Lenslet array samples the image plane and the void created due to demagnification is used for spectral dispersion

An image of a 2D field formed by the telescope or an imaging lens is sampled by a set of lenslets (Figure 1.3). The image sampled by the lenslet gets shrunk or demagnified and occupies a small region with a void surrounding it. Each of the demagnified images is collimated and spectrally dispersed using a grating. The dispersed spectra when re-imaged forms a spectral streak of the demagnified image thus giving wavelength information at multiple points on the 2D field.

### 1.3.3 Snapshot Spectroscopy with a fiber bundle

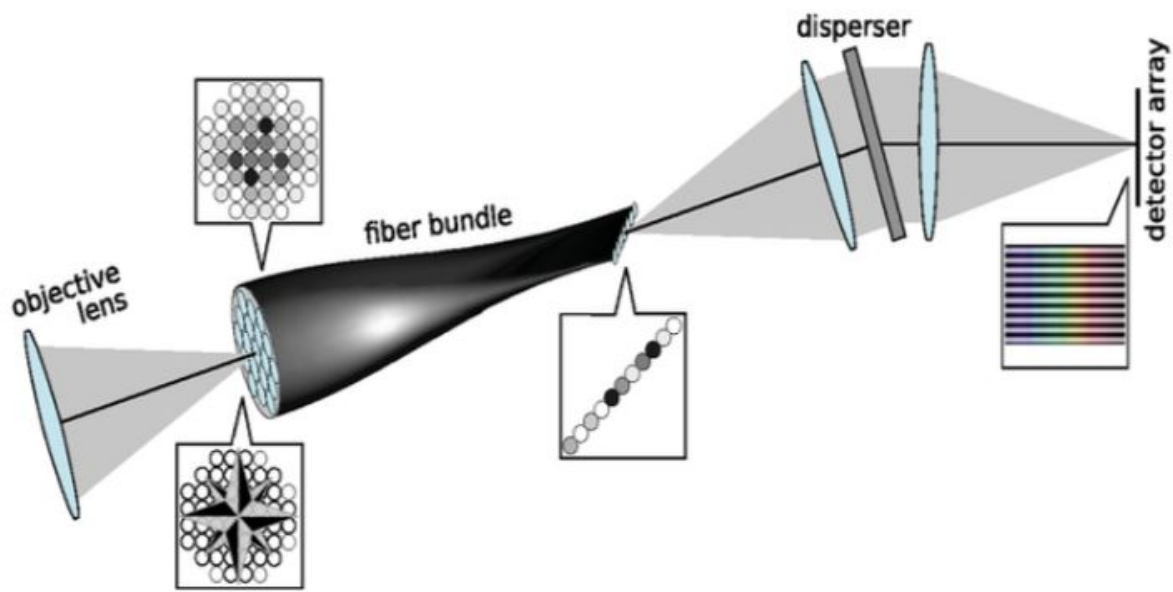

Fig 1.4 A fiber bundle samples a 2D image plane and re-formats it into 1D slit which is dispersed using a spectrograph

An optic fiber bundle is placed at the image plane of the telescope (Figure 1.4). The fibers are oriented in such a way that the 2D field sampled by the fibers are converted to 1D slit structure. A grating is used to obtain the spectra of the 1D slit. As we obtain the spectral information of every point on the 1D slit, we effectively have the spectra of the 2D image forming a hyperspectral image.

All the techniques mentioned above sample the image plane and uses a grating based spectrograph to obtain capabilities. A novel instrument technique is developed by sampling the pupil plane instead of the image plane and uses a Fabry-Perot interferometer as a spectral disperser.

## 1.5 Structure of this Thesis

Each chapter is independent of the contents from the other chapters and is self-sustained. As the chapters talk about instruments or designs, the conclusion of the chapters also contains possible improvements and future work of those instruments. The second chapter discusses the Zeemax design for MAST and DKIST. It also talks about tolerances considered while designing the achromats. The third chapter discusses the snapshot spectroscopy technique demonstrated at MAST and the data reduction technique. It also discusses the snapshot spectro-polarimetry experiment done at IIA and presents its results.

# 2 Snapshot Spectrograph Designs

## 2.1 Optical design for pupil sampling Snapshot Spectrograph

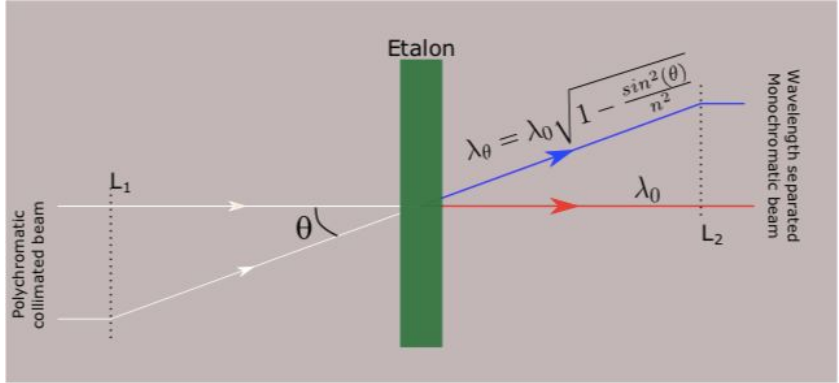

Fig. 2.1 Blue shift in transmitted wavelength for a ray traveling at an angle with respect to the ray traveling normal to the etalon

A schematic of the optical setup is displayed in Fig. 2.1. As shown in this figure the field of view (FOV) is limited using a field stop. The beam exiting the field stop is collimated using a lens L1 and the pupil image is formed behind L1 at a distance equal to its focal length. This pupil plane is sampled by a lenslet array to get individual identical images of the region of interest (ROI) at the focal plane of the lenslet (Shack Hartmann Image Plane). The lenslet images are re-collimated using lens L2, which is analogous to lens L1 in Fig. 2.2, and the beam is made to pass through an etalon at angles which are determined by the focal length of the lens L2 (f) and the distance of the object (d) from the optic axis according to the relation,

$$\theta = \tan^{-1}\left(\frac{d}{f}\right). \qquad (2.1)$$

As the distance from the optic axis increases, the peak transmitted wavelength shift towards the bluer side. Since the system is symmetric along the radial direction, all points at the same radial distance from the optic axis are at the same wavelength. The wavelength decreases radially according to the relation,

$$\Delta\lambda = \left(\frac{\lambda_0}{2}\right) \times \left(\frac{\theta}{\mu}\right)^2. \qquad (2.2)$$

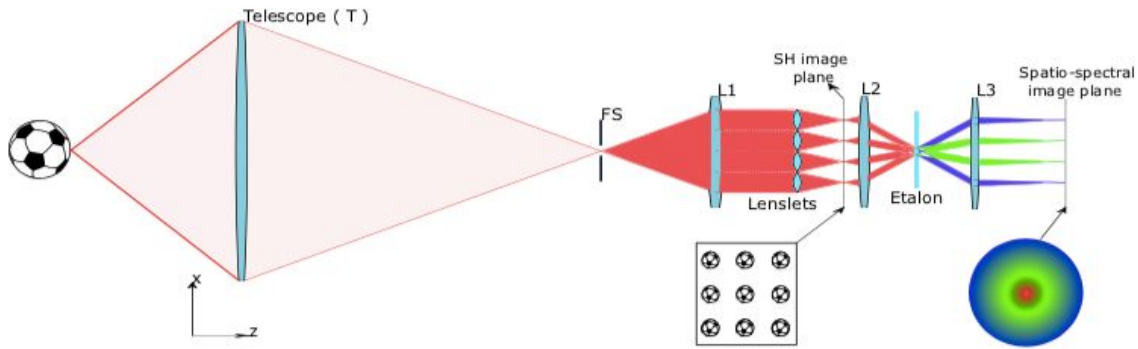

Fig. 2.2 Optical design for pupil sampling snapshot spectrograph

The wavelength sampled pupil is imaged using lens L3, which is analogous to lens L2 in Fig. 2.2, to get multiple images of the region of interest onto the spatio-spectral image plane. Hence, we get multiple images of ROI sampled at different wavelengths to get a hyperspectral image of the object at the spatio-spectral image plane.

## 2.2 50 cm Telescope at MAST

The MAST telescope is an off-axis Gregorian type with 50 cm aperture. The primary mirror (M1) is made of Zerodur and has a focal length of 2 m (focal number = 4). Fig. 2.3 shows the optical layout (a) and the mechanical design (b) of the telescope. The field stop placed at prime focus allows 6 arcmin field-of-view to pass through the remaining optics. The secondary mirror (M2) placed close to the field stop collimates the light. The Coude train, which consists of three mirrors (M3, M4, and M5), sends the beam vertically down to the observing room. The telescope is placed on an alt-azimuth mount and thus causes image rotation. The beam is sent through an image derotator (M6, M7, and M8) before reaching the folding mirror M9. From M9 the collimated beam is sent to the back-end instruments in the observing room.

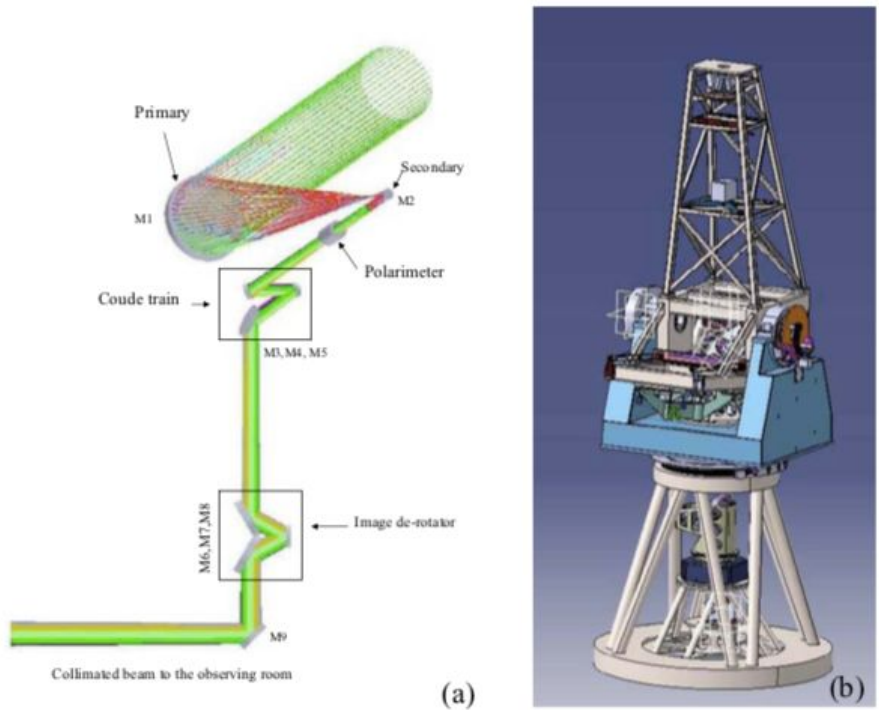

Fig 2.3 Optical layout (a) and the mechanical design (b) of the MAST telescope (courtesy: AMOS, Belgium).

## 2.3 Snapshot Spectrograph for Multi Application Solar Telescope (MAST)

The following off-the-shelf elements are used in the pupil sampling spectrograph design for MAST.

Table 2.1 Optical Design Specification of Elements in Snapshot Spectrograph for MAST

| Element | Specification | Vendor and Model |
|---|---|---|
| Telescope (T) Aperture | 90 mm | Edmund Optics #54-567 |
| Telescope (T) Focal Length | 849.9 mm | |
| Field Stop | 3.858 mm | |
| L1 Focal Length | 175 mm | Edmund Optics #49-363 |

| SH Lenslet Pitch | 1 mm | |
| --- | --- | --- |
| SH Lenslet Focal Length | 45 mm | |
| L2 Focal Length | 200 mm | Edmund Optics #88-596 |
| Fabry Perot Type | Air-Spaced Etalon | |
| FP FSR and FWHM | 3.75 Å and 235 mÅ | |
| L3 Focal Length | 150 mm | Edmund Optics #49-391 |
| FOV | 1.5 arcmin | |

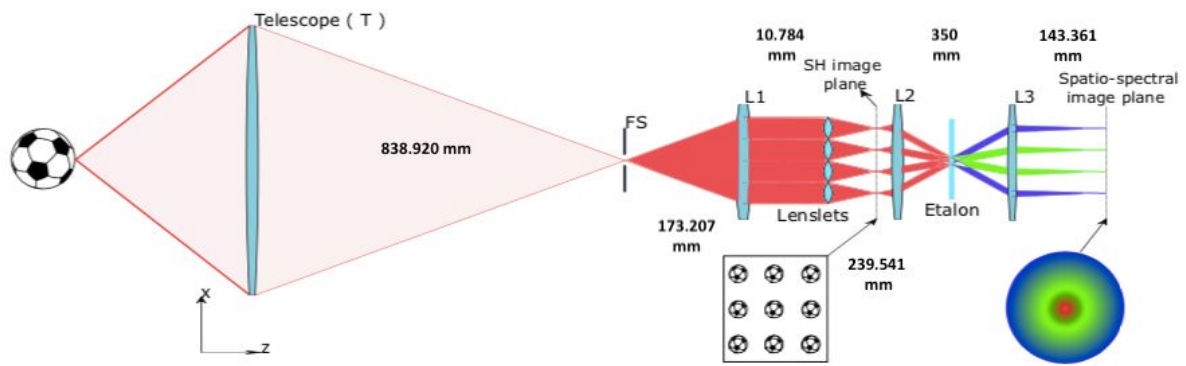

Fig 2.4 Pupil Sampling Snapshot Spectrograph Design for MAST

## 2.3.1 Point Spread Function

The point spread function (PSF) describes the response of an imaging system to a point source or point object. A more general term for the PSF is a system's impulse response, the PSF being the impulse response of a focused optical system.

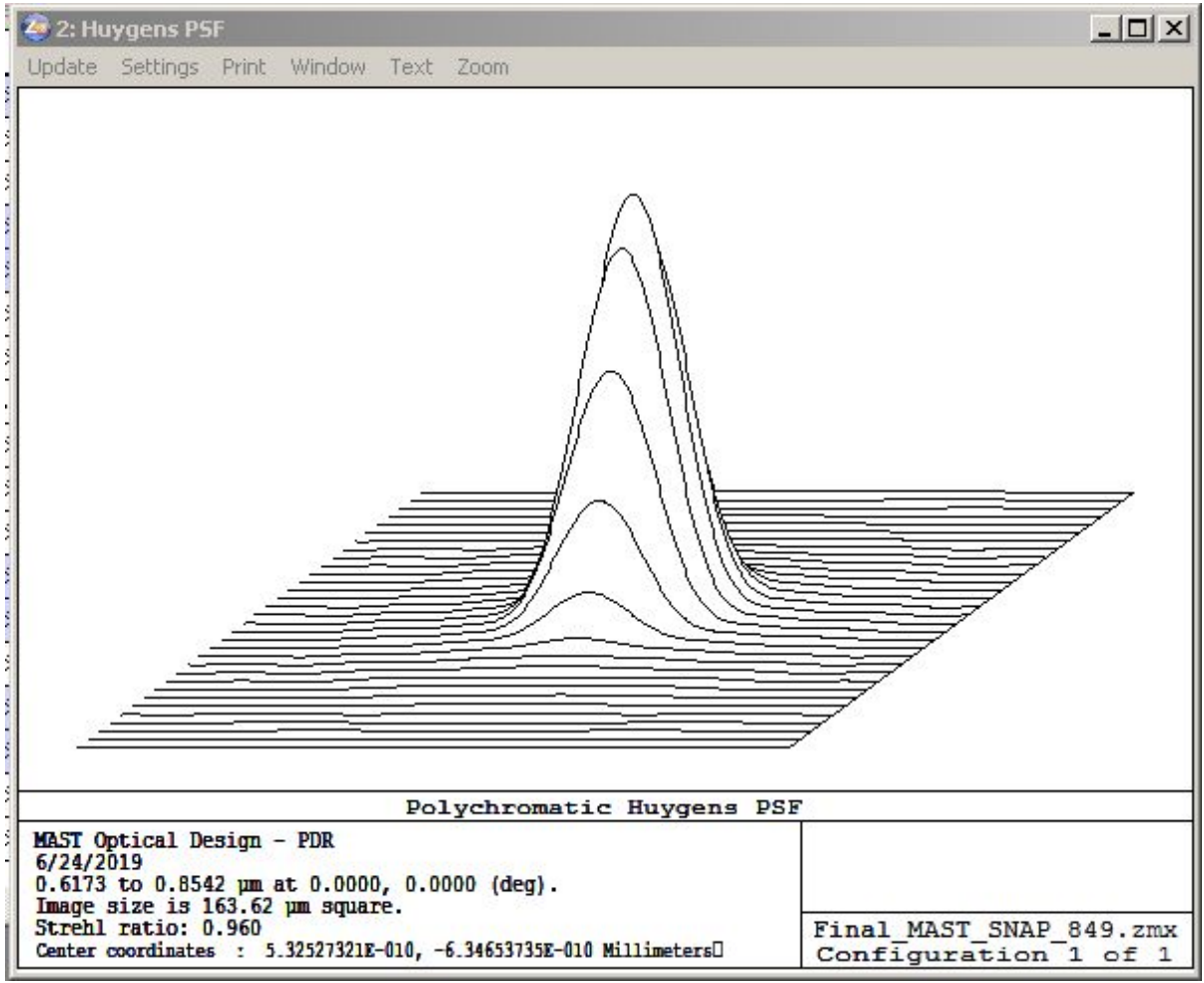

Fig 2.5 PSF of the Snapshot Spectrograph for MAST

Figure 2.5 displays the point spread function (PSF) for the pupil sampling spectrograph for MAST. The Strehl ratio is 0.960.

### 2.3.2 RMS Wavefront Error

The root-mean-square wavefront error or RMS wavefront error is a specific average over the wavefront phase errors in the exit pupil.

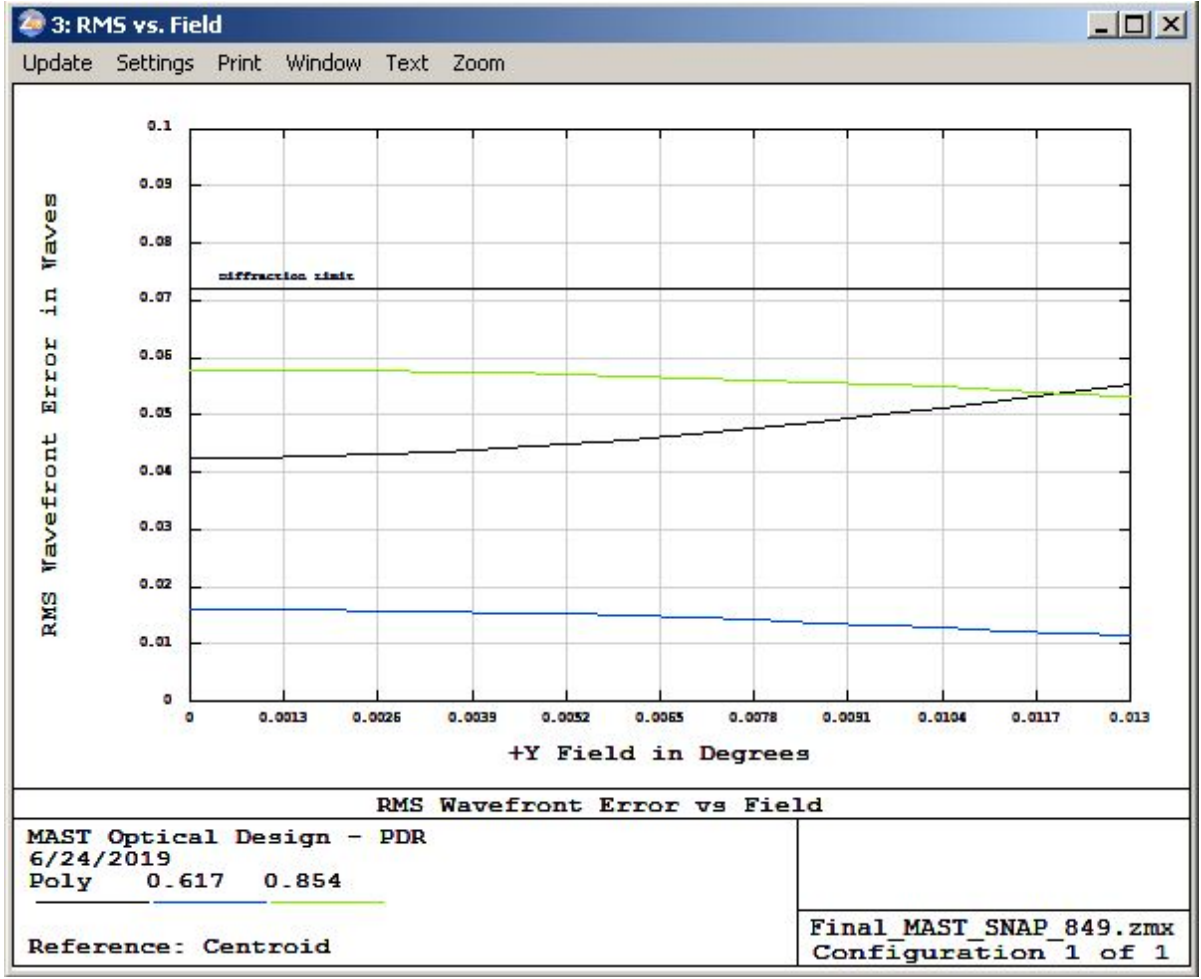

Fig 2.6 RMS Wavefront error for different wavelength and fields for MAST

Figure 2.6 displays the RMS wavefront error over the full field for MAST. We can see, it is below the diffraction limit.

### 2.3.3 Spot Diagram

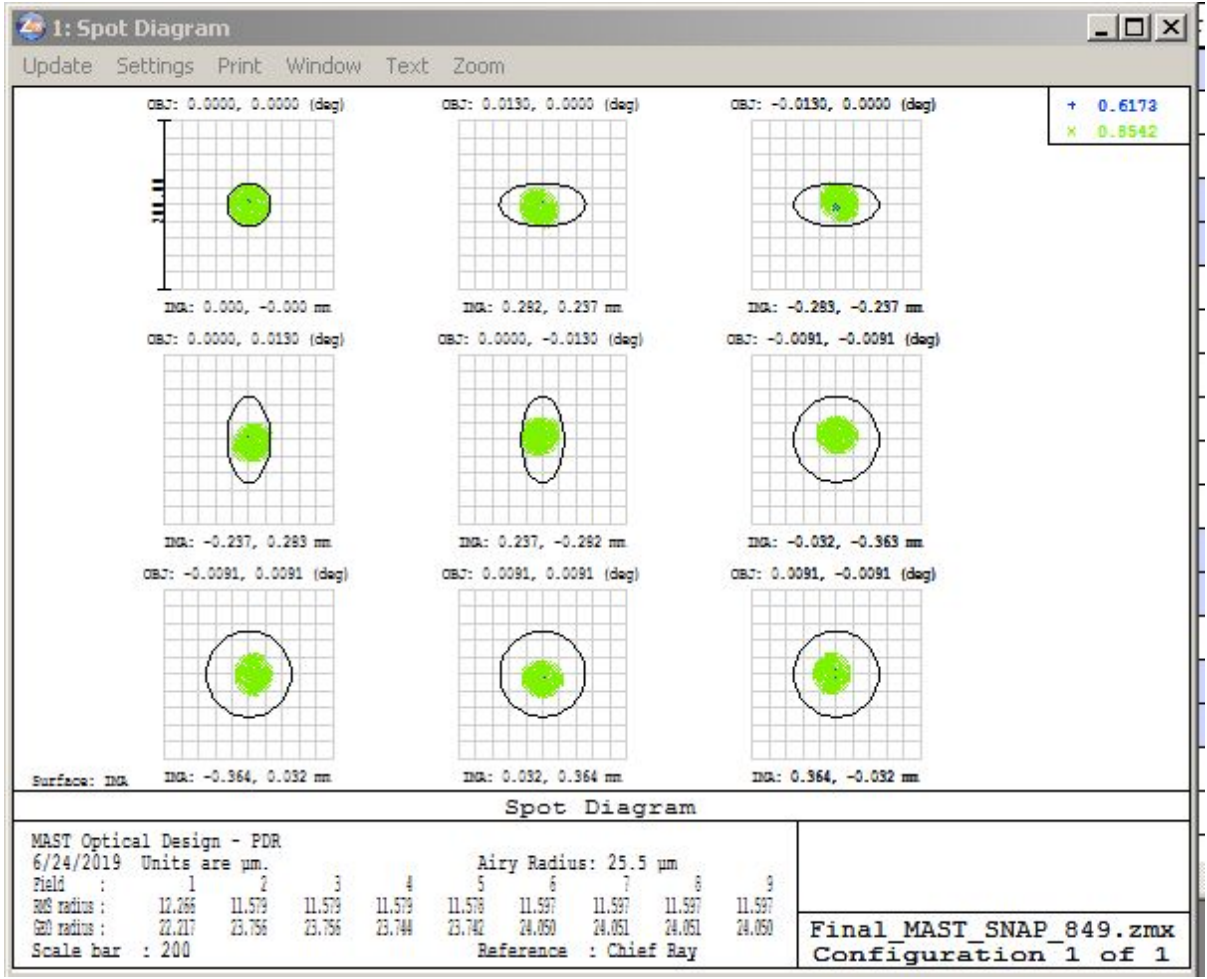

Fig 2.7 Spot Diagram for different wavelength and fields for MAST

Figure 2.7 displays the spot diagram over wavelengths 6302 Å and 8542 Å over different fields. We can see the maximum spot radius is 24.05 μm which is below the airy disc radius 25.5 μm.

### 2.3.4 Design Parameters for Snapshot Spectrograph for MAST

Table 2.2 Parameters and values for MAST Snapshot Spectrograph

| Parameters | Values |
| --- | --- |
| Airy Disc Radius | 25.5 μm |
| Spot Size | 24.05 μm |
| Strehl Ratio | 0.960 |
| RMS Wavefront Error | 0.06 waves |

## 2.3.5 Angle Distribution at image plane due to FP for MAST

We have calculated angles at a 1k X 1k detector with 6.4 μm pixel size using the lenslet array as mentioned above.

We have assumed the detector axis is aligned with the principal axis of the system and the Fabry Perot. We have calculated angles at each pixel from the line joining the center of the imaging lens and the pixel with the principal axis. To calculate the microlens array mask, as given in Table 2.1, we have used a 1 mm pitch microlens array, 200 mm focal length collimating lens and 150 mm focal length imaging lens. We infer that because the system is centrally symmetric, at same radial distances from the detector center, the angles will be the same, and Hence according to equation 2.2, all pixels at the same radial distance will be at the same wavelength. We also note that the maximum angular variation is 1.4 degrees. Figure 2.8 shows the angle variation on the detector with the microlens array masks.

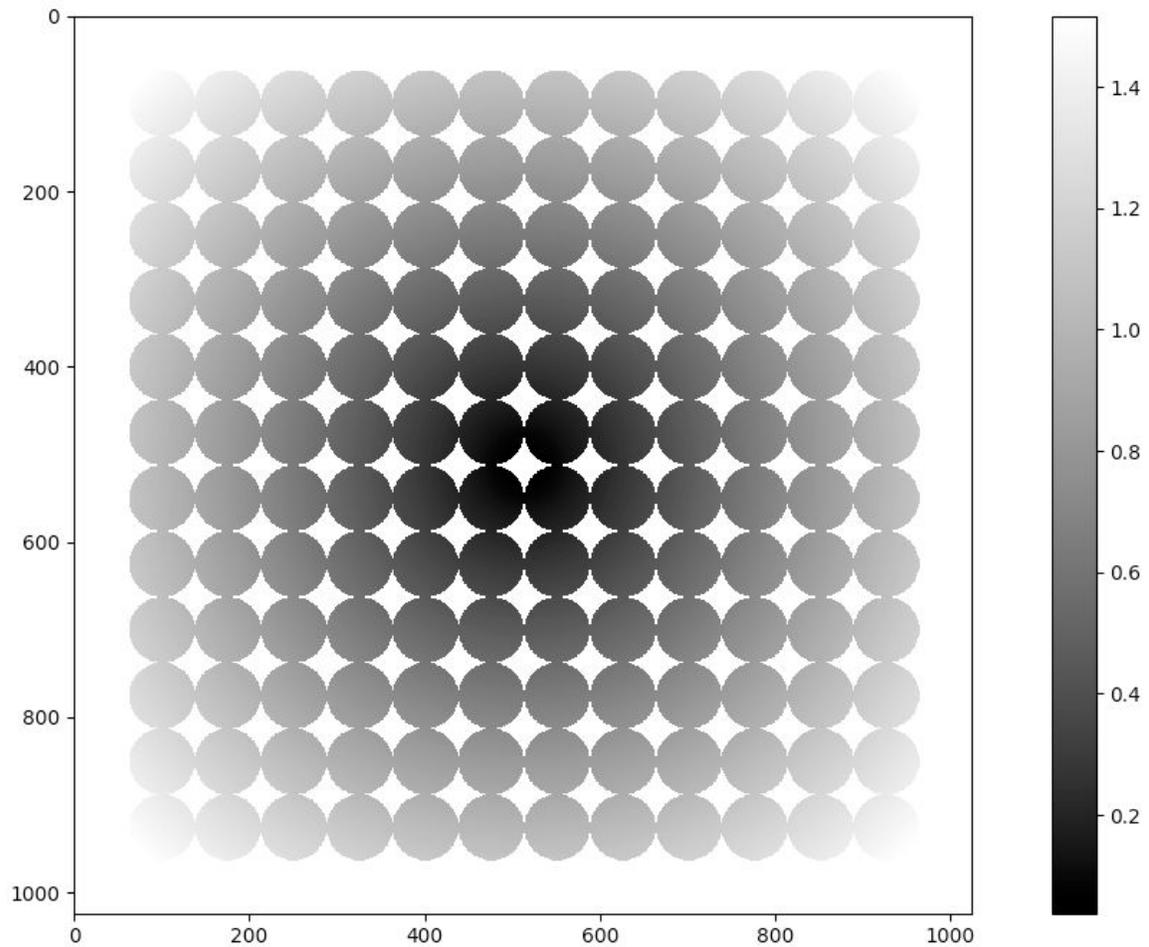

Fig 2.8 Angles in degrees on the Detector when 1k x 1k detector with 6.4 μm pixel size is used with microlens arrays with 1 mm pitch.

## 2.4 Daniel K. Inouye Solar Telescope (DKIST)

The DKIST is an off-axis Gregorian telescope (Hubbard, Robert [3]) with 4240 mm diameter primary mirror with aperture stop of 4000 mm with the radius of curvature 16000 mm. The secondary mirror has a radius of curvature of 2081.255 mm. These are followed by Coudé transfer optics which consists of a fold mirror and a relay mirror with a radius of curvature (ROC) 5979.243 mm. These are followed by mount base transfer optics (fold mirrors) and finally to Coudé Rotator optics which consists of fold mirror, a collimator (ROC 21442.354 mm) and a coma corrector and deformable mirror. (Figure 2.9, 2.10 and 2.11)

Table 2.3 Optical Prescription for different mirror surfaces of DKIST

| Element | | Conic Constant | Radius of Curvature (mm) | Curve |
|---|---|---|---|---|
| **OSS Gregorian Optics** | | | | |
| M1 | Primary | -1 | 16,000 | Concave |
| M2 | Secondary/FSM | -0.5393103 | 2,081.255 | Concave |
| **OSS Coudé Transfer Optics** | | | | |
| M3 | Fold Mirror | – | – | Flat |
| M4 | Relay Mirror | -0.3717299 | 5,979.243 | Concave |
| **Mount Base Assembly Coudé Transfer Optics** | | | | |
| M5 | FSM | – | – | Flat |
| M6 | Fold Mirror | – | – | Flat |
| **Coudé Rotator Optics** | | | | |
| M7 | Fold Mirror | – | – | Flat |
| M8 | Collimator | -1 | 21,442.354 | Concave |
| M9 | Coma Corrector | – | – | Flat w/coma[1] |
| M10 | Deformable Mirror | – | – | Flat |
| BS1 | WFC Beam Splitter | – | – | Flat |

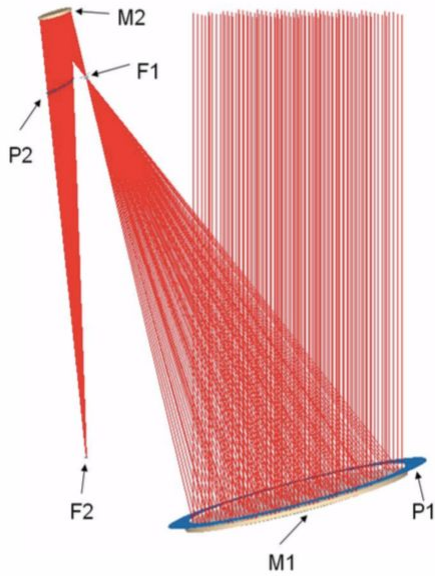

Fig 2.9 Gregorian Optics of DKIST (Hubbard, Robert [3])

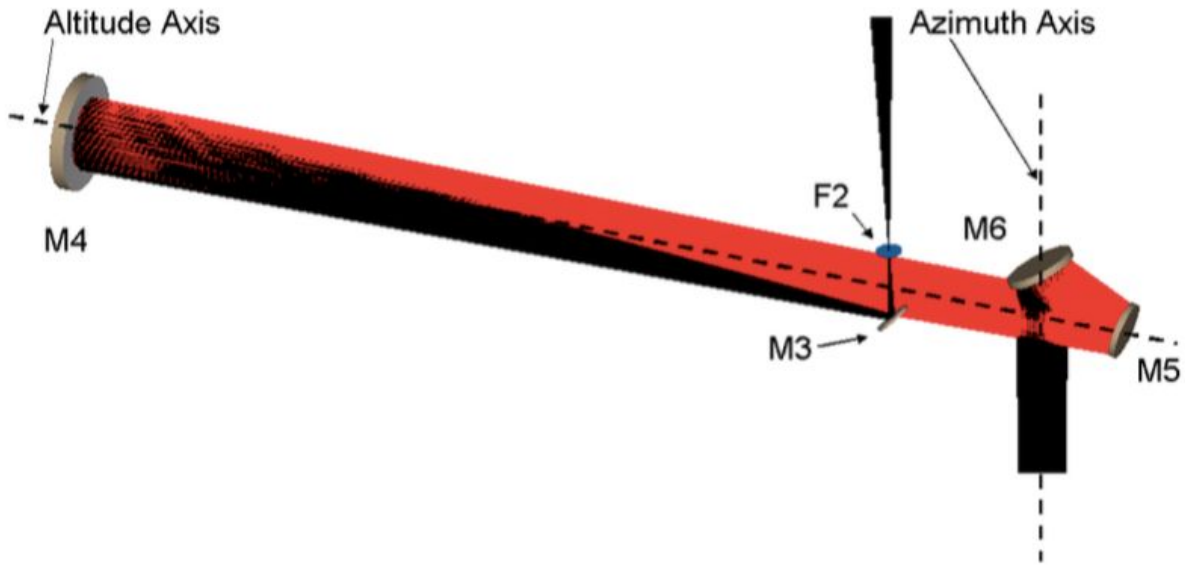

Fig 2.10 The Coudé Transfer Optics include two groups of components. The OSS group (M3 and M4) move with the Optical Support Structure as it rotates about the altitude axis. The other group consists of M5 and M6, which are fixed to the Mount Base Assembly. (Hubbard, Robert [3])

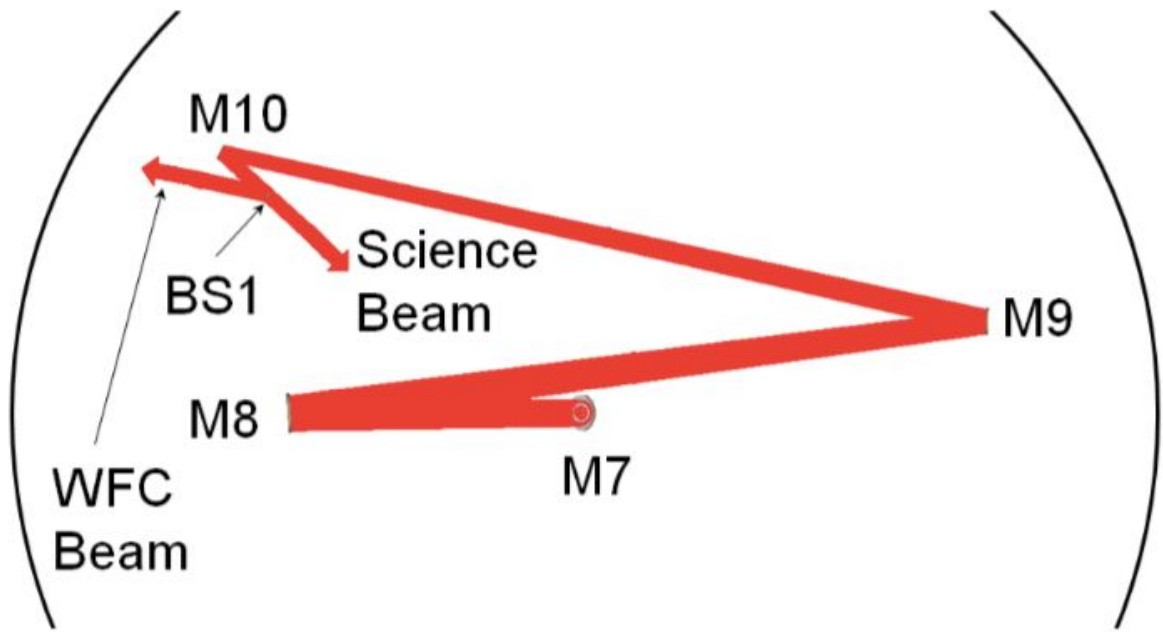

Fig 2.11 The Coudé Rotator Optics include M7 through M10 (the deformable mirror) and BS1, the beam splitter that reflects a small fraction of the beam into the wavefront correction system. (Hubbard, Robert [3])

## 2.5 Snapshot Spectrograph design for DKIST

We have designed four achromats, two for collimation(L1, L2) and two for imaging (T, L3) and selected an off-the-shelf microlens array. We have also optimized the distances between various components and have done the tolerance analysis to measure the tolerances over different parameters like the radius of curvature, tilts, distances between components and material purity (Abbe numbers). We are presenting an optical prescription for the achromats, the distance between components, worst and best case performance parameters like Spot Diagram, PSF and the RMS wavefront error.

### 2.5.1 Specifications

Table 2.4 Optical Design Specifications of the elements of snapshot spectrograph for DKIST

| Element | Specification |
| --- | --- |
| Telescope (T) Aperture | 280 mm |
| Telescope (T) Focal Length | 3365 mm |
| Field Stop | 14.026 mm |
| L1 Focal Length | 175 mm |
| SH Lenslet Pitch | 1.5 mm |
| SH Lenslet Focal Length | 24.3 mm |
| L2 Focal Length | 300 mm |
| Fabry Perot Type | Air-Spaced Etalon |
| FP FSR and FWHM | 5 Å and 235 mÅ |
| L3 Focal Length | 300 mm |
| FOV | 30 arcsec |

## 2.5.2 Achromatic Doublet Designs

### 2.5.2.1 First Imaging Lens (T)

Table 2.5 Optical Prescription of the first imaging lens (T) for DKIST

| Element | Value |
| --- | --- |
| The radius of Curvature 1st Surface | 1706.649 mm |
| The Radius of Curvature 2nd Surface | -838.721 mm |
| Thickness | 40 mm |
| Material | N-PSK57 |
| Air Gap | 5 mm |
| The radius of Curvature 1st Surface | -840.782 mm |
| The radius of Curvature 2nd Surface | Infinity |
| Thickness | 40 mm |
| Material | BAF9 |
| Focal Length | 3365 mm |
| Semi Diameter | 140 mm |

### 2.5.2.2 Collimating Lens (L1)

Table 2.6 Optical Prescription of the collimating (L1) lens for DKIST

| Element | Value |
| --- | --- |
| The radius of Curvature 1st Surface | 174.652 mm |
| The Radius of Curvature 2nd Surface | -69.191 mm |

| Thickness | 10 mm |
|---|---|
| Material | P-PK53 |
| Air Gap | 0 mm |
| The radius of Curvature 1st Surface | -69.191 mm |
| The radius of Curvature 2nd Surface | -135.658 mm |
| Thickness | 4 mm |
| Material | BAF9 |
| Focal Length | 175 mm |
| Semi Diameter | 7.506 mm |

2.5.2.3 Micro Lens Array

Table 2.7 Optical Design Specification of the microlens array for DKIST

| Element | Value |
|---|---|
| Lenslet Model | Oko Tech APO-GB-P1500-F24.3 (633) |
| Pitch | 1.5 mm |
| Focal Length | 24.3 mm |

2.5.2.4 Collimating Lens (L2)

Table 2.8 Optical Prescription of the collimating (L2) lens for DKIST

| Element | Value |
|---|---|
| The radius of Curvature 1st Surface | 331.023 mm |

| The radius of Curvature 2nd Surface | -81.369 mm |
|---|---|
| Thickness | 13.59 mm |
| Material | BAK50 |
| Air Gap | 0 mm |
| The radius of Curvature 1st Surface | -81.369 mm |
| The Radius of Curvature 2nd Surface | -231.877 mm |
| Thickness | 6 mm |
| Material | N-SF5 |
| Focal Length | 300 mm |
| Semi Diameter | 21.831 mm |

2.5.2.5 Imaging Lens (L3)

Table 2.9 Optical Prescription of the imaging (L3) lens for DKIST

| Element | Value |
|---|---|
| The radius of Curvature 1st Surface | 117.218 mm |
| The radius of Curvature 2nd Surface | -137.987 mm |
| Thickness | 13.59 mm |
| Material | N-BK7 |
| Air Gap | 0 mm |
| The radius of Curvature 1st Surface | -137.987 mm |
| The radius of Curvature 2nd Surface | Infinity |
| Thickness | 6 mm |

| Material | N-SF5 |
|---|---|
| Focal Length | 300 mm |
| Semi Diameter | 20.006 mm |

## 2.5.3 Pupil Sampling Snapshot Spectrograph Design for DKIST

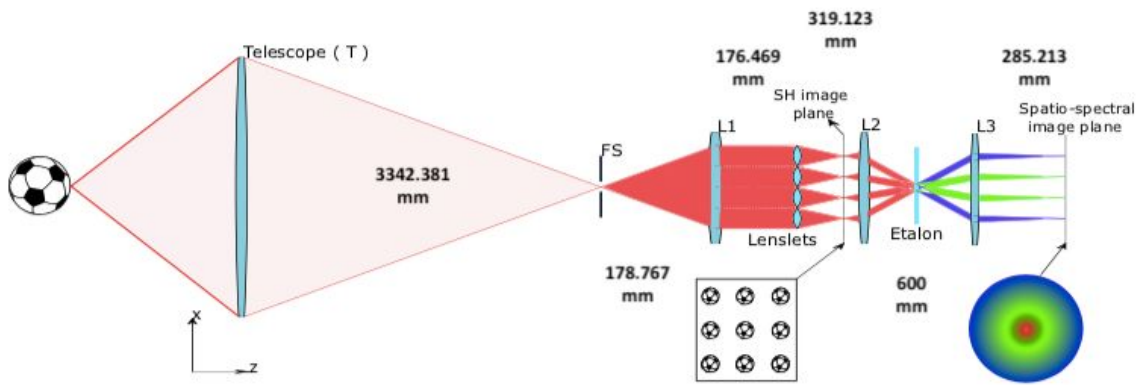

Fig 2.12 Optical design of the Snapshot Spectrograph for DKIST

Fig 2.12 displays the different components of the pupil sampling spectrograph with the distances between them.

## 2.5.4 Tolerance Analysis

The following tables (Table 2.10 and Table 2.11) present tolerances for various parameters in optical prescriptions of lenses and the design of snapshot spectrograph.

Table 2.10 Tolerance data for lenses for DKIST

| Element | Tolerances |
|---|---|
| Radius of Curvature | +/- 0.125 mm |
| Thickness | +/- 0.05 mm |

| Tilt | +/- 0.05 degrees |

Table 2.11 Abbe Tolerances of the material at different surfaces.

| Abbe Number | Min | Max |
|---|---|---|
| 68.399 | -0.171 | 0.171 |
| 47.958 | -0.120 | 0.120 |
| 66.220 | -0.166 | 0.166 |
| 28.533 | -0.071 | 0.071 |
| 57.990 | -0.145 | 0.145 |
| 32.251 | -0.081 | 0.081 |
| 64.167 | -0.160 | 0.160 |
| 32.251 | -0.081 | 0.081 |

Appendix A has detailed tolerance analysis results.

We have considered variation in point spread function (PSF), the Strehl ratio, spot radius, and RMS wavefront error to judge a Monte-Carlo simulation performance. Table 2.12 summarises the worst case and best case values of the above parameters.

2.5.4.1 Worst Case Point Spread Function (PSF)

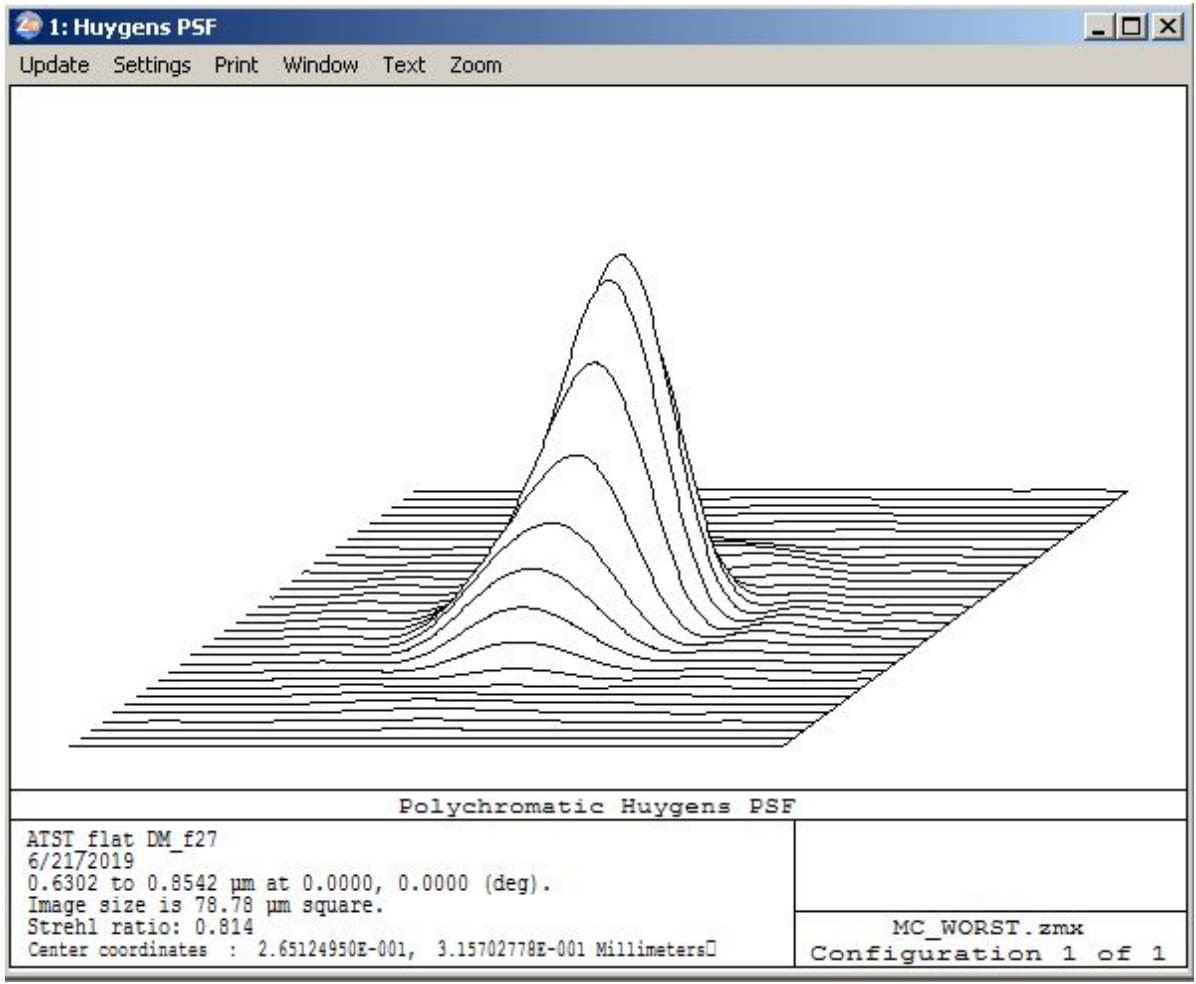

Fig 2.13 Worst case PSF during Monte Carlo Simulations for DKIST

Figure 2.13 displays PSF of the worst PSF attained during the Monte-Carlo simulations. The Strehl ratio is degraded to 0.814 from 0.981.

2.5.4.2 Worst Case Spot Diagram

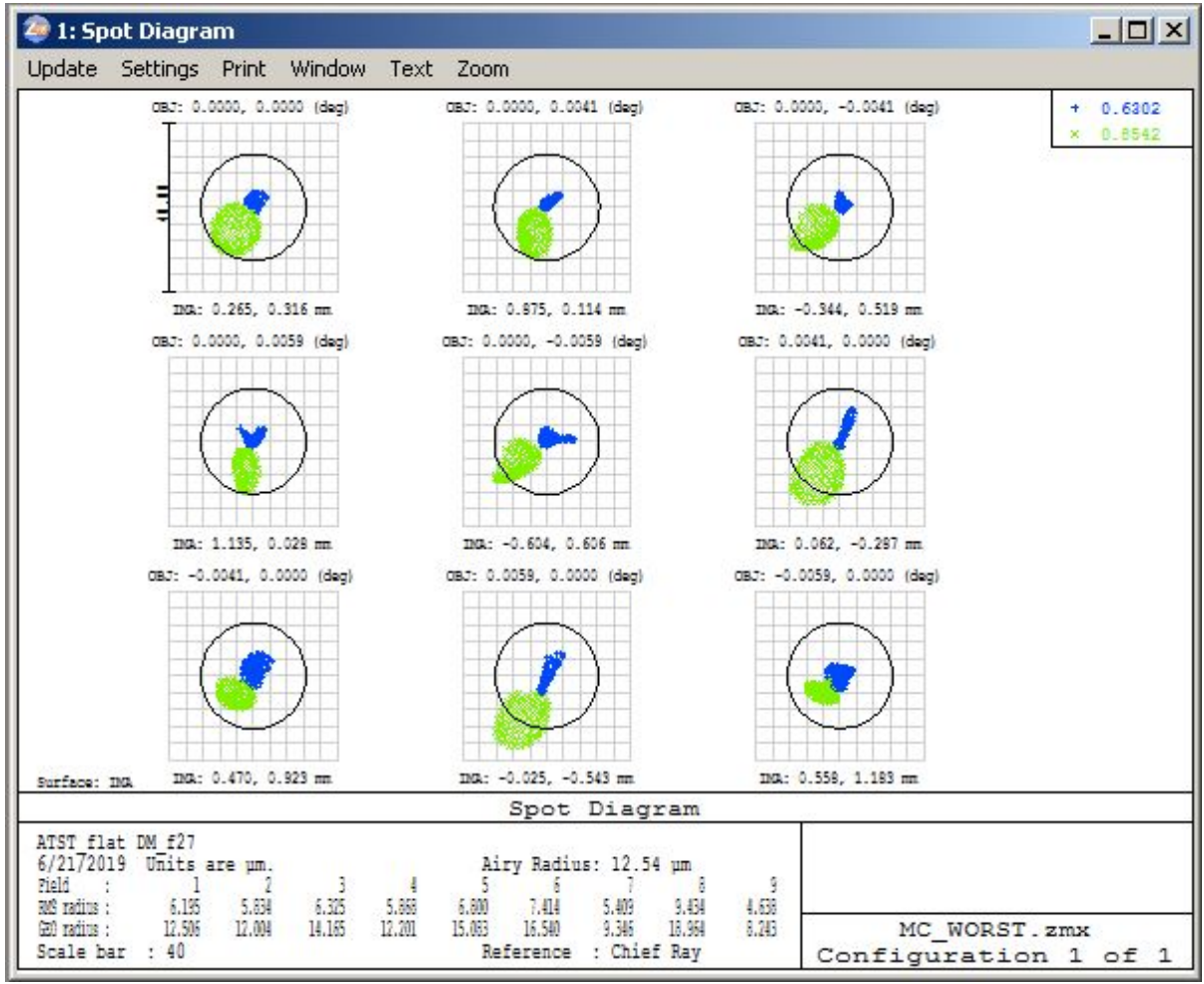

Fig 2.14 Worst case Spot Diagram during Monte Carlo Simulations for DKIST

Figure 2.14 displays Spot Diagram of the worst spot diagram attained during the Monte-Carlo simulations. The spot radius is degraded to 18 μm from 7 μm.

2.5.4.3 Worst Case RMS Wavefront Error

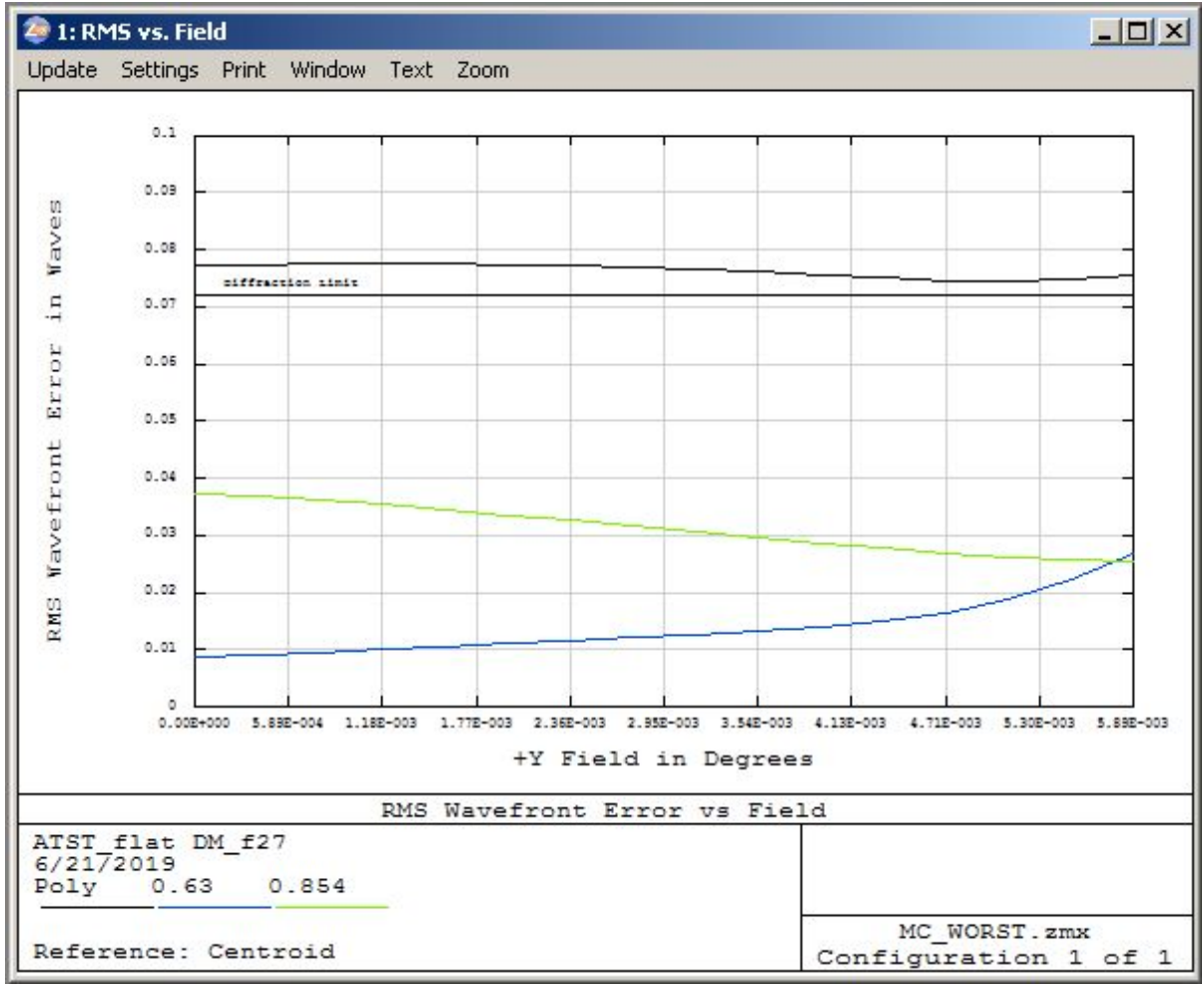

Fig 2.15 Worst case RMS Wavefront Error during Monte Carlo Simulations for DKIST

Figure 2.15 displays RMS wavefront error of the worst RMS wavefront error attained during the Monte-Carlo simulations. The RMS wavefront error is degraded to 0.08 waved from 0.04 waves.

2.5.4.4 Best Case Point Spread Function (PSF)

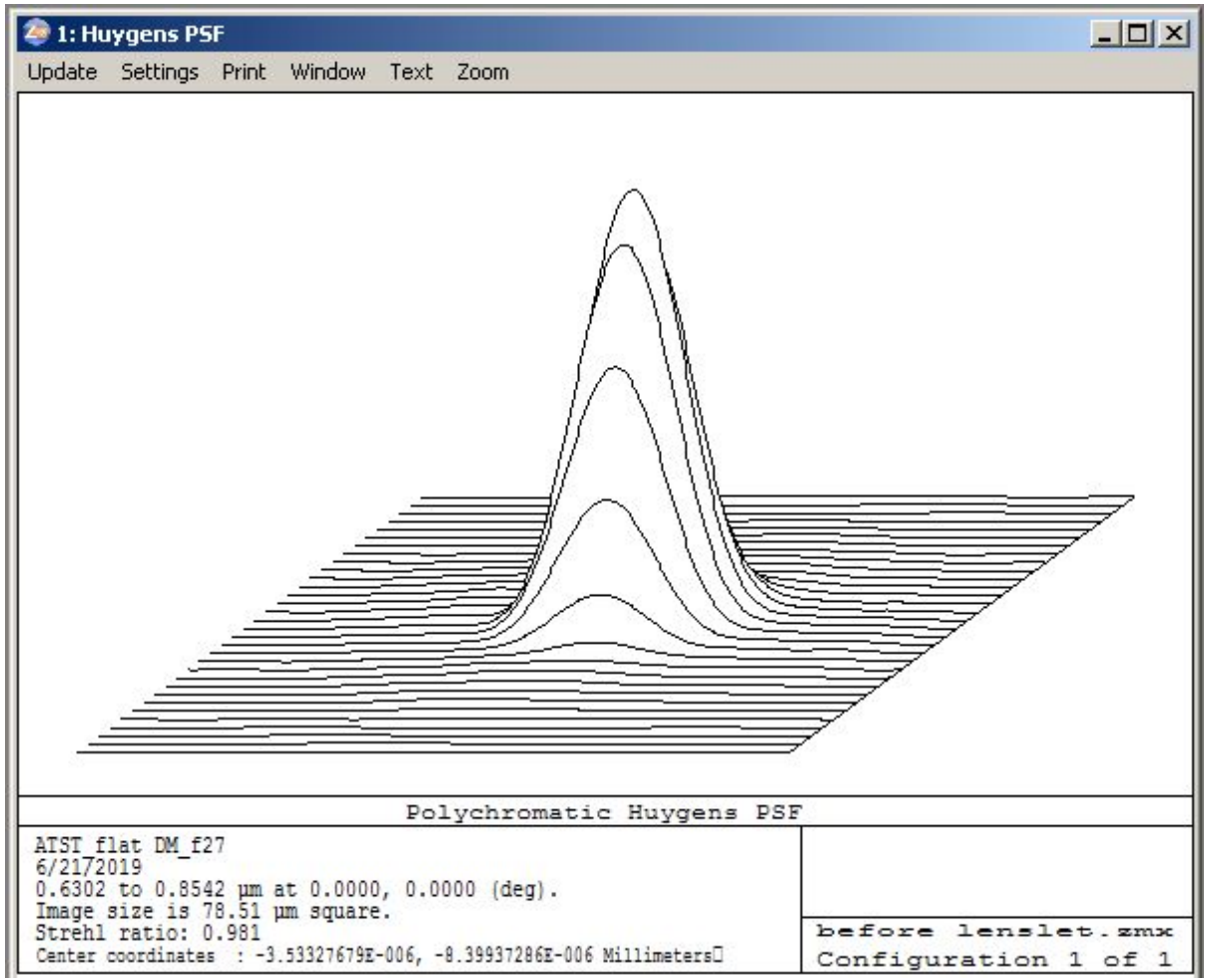

Fig 2.16 Best case PSF for DKIST

Figure 2.16 displaying best PSF attained with the Strehl ratio 0.981.

2.5.4.5 Best Case Spot Diagram

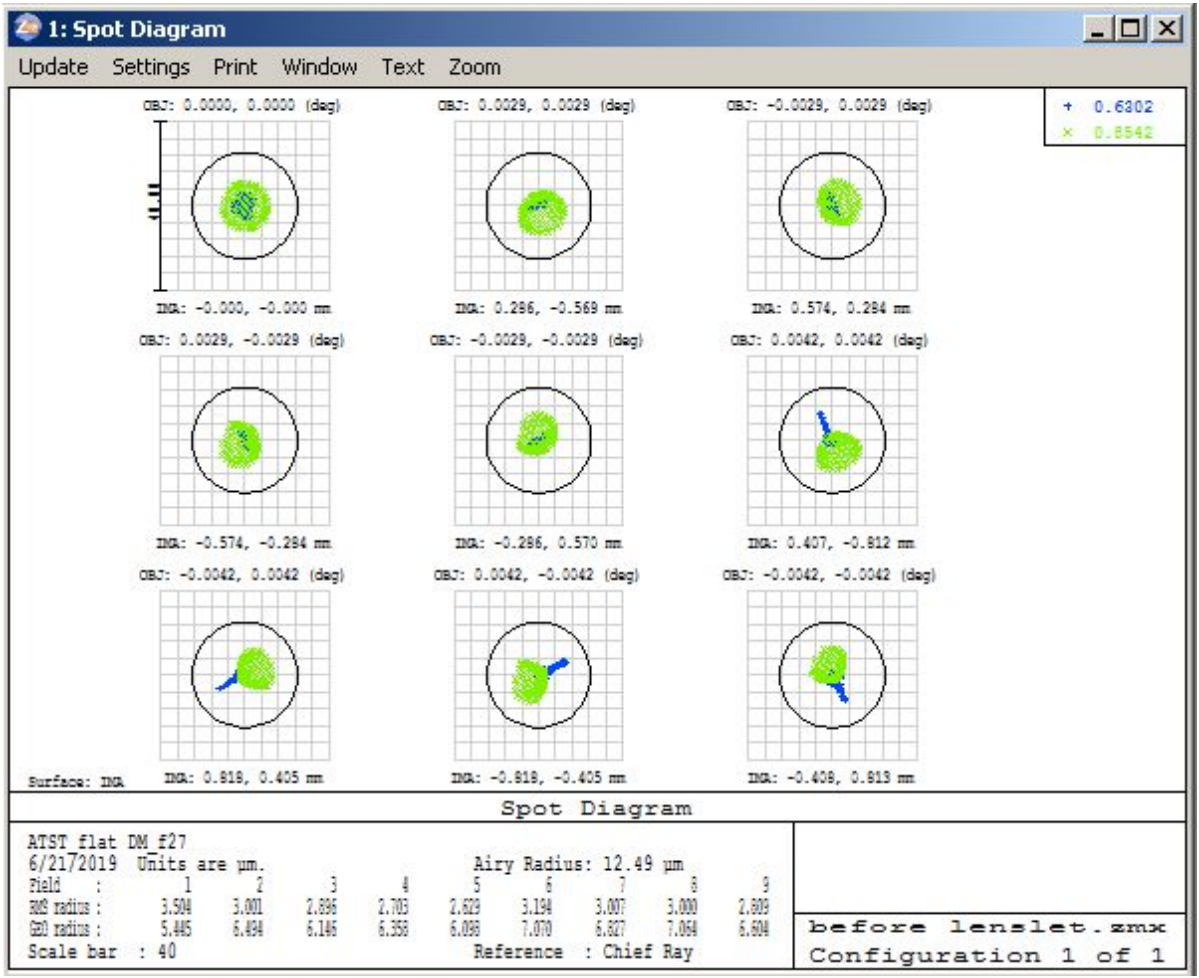

Fig 2.17 Best case Spot Diagram for DKIST

Figure 2.17 displaying best Spot radius attained with the spot radius 7.07 μm.

2.5.4.6 Best Case RMS Wavefront Error

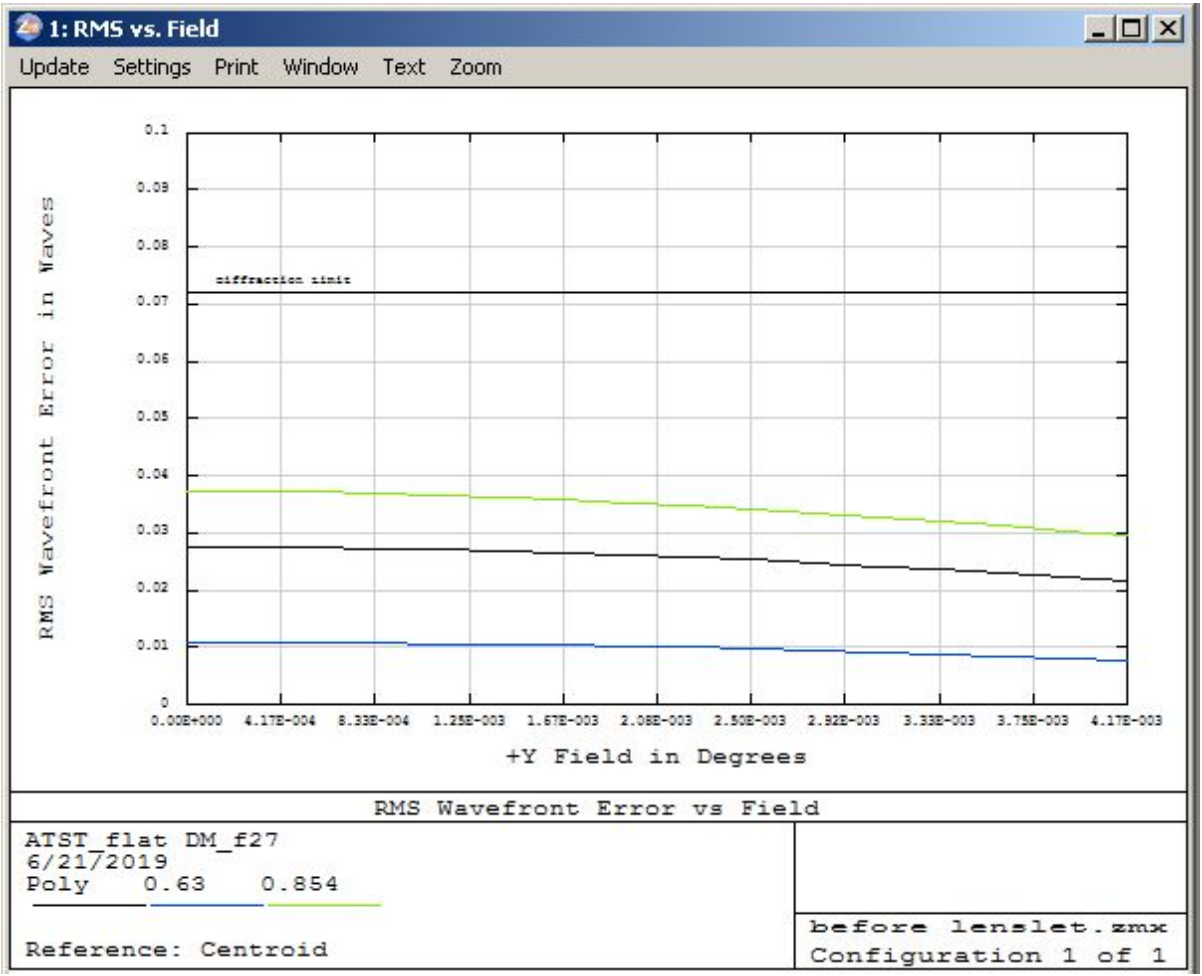

Fig 2.18 Best case RMS Wavefront Error for DKIST

Figure 2.18 displaying best RMS wavefront error attained with the RMS wavefront error 0.04 waves.

## 2.5.5 Design Parameters for Snapshot Spectrograph for DKIST

Table 2.12 Parameters and values for DKIST Spectrograph

| Parameter | Worst Case Value | Best Case Value |
|---|---|---|
| Airy Disc Radius | 12.54 μm ||
| Spot Size | 18 μm | 7 μm |
| Strehl Ratio | 0.814 | 0.981 |
| RMS Wavefront Error | 0.08 waves | 0.04 waves |

## 2.5.6 Angle Distribution at image plane due to FP for DKIST

We have calculated angles at a 2k X 2k detector with 6.4 μm pixel size using the lenslet array as mentioned above.

Similar to the work we did for MAST, We have assumed the detector axis is aligned with the principal axis of the system and the Fabry Perot. We have calculated angles at each pixel from the line joining the center of the imaging lens and the pixel with the principal axis. To calculate the microlens array mask, as given in Table 2.4, we have used a 1.5 mm pitch microlens array, 300 mm focal length collimating lens and 300 mm focal length imaging lens. We infer that because the system is centrally symmetric, at same radial distances from the detector center, the angles will be the same, and Hence according to equation 2.2, all pixels at the same radial distance will be at the same wavelength. We also note that the maximum angular variation is 1.4 degrees. Figure 2.19 shows the angle variation on the detector with the microlens array masks.

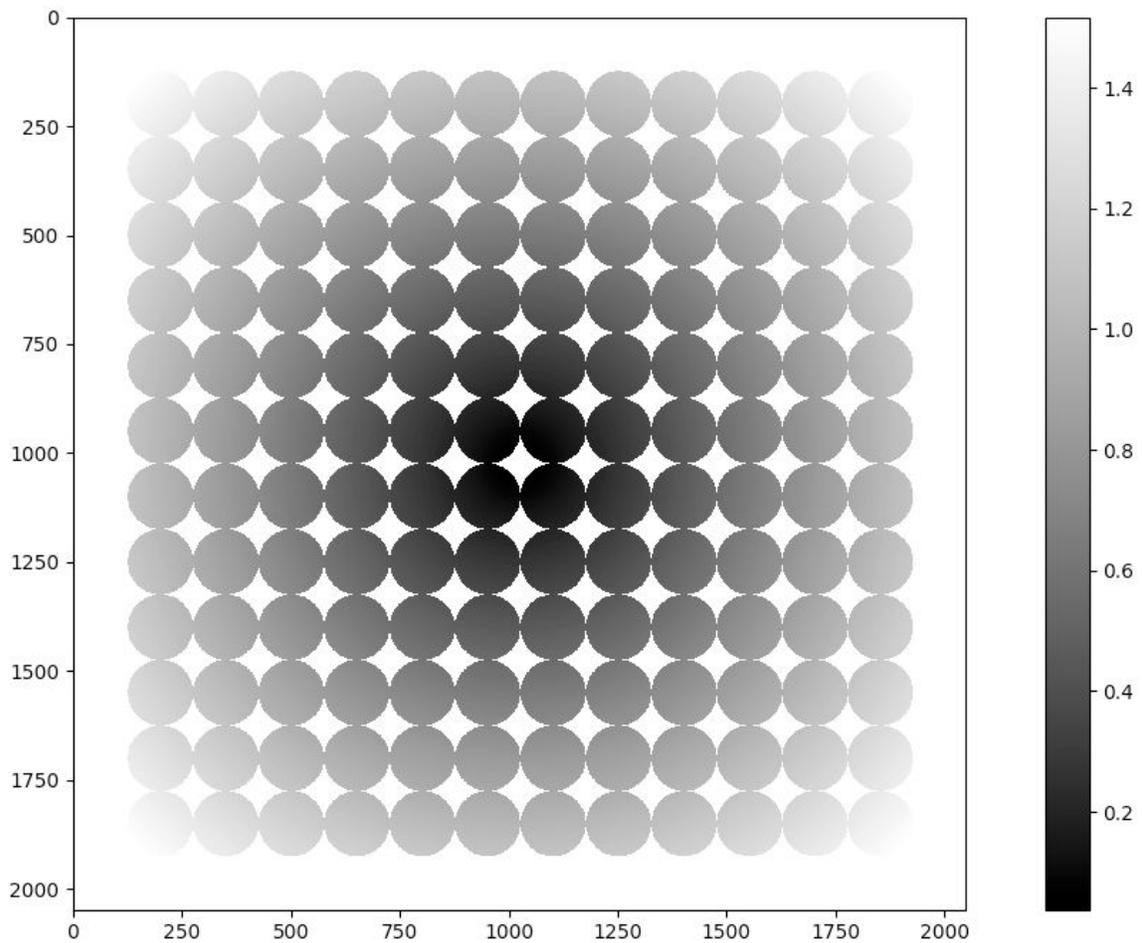

Fig 2.19 Angles in degrees on the Detector when 2k x 2k detector with 6.4 μm pixel size is used with microlens arrays with 1.5 mm pitch

## 2.6 Summary and Future Work

We have designed spectrograph designs for MAST and DKIST. For MAST we have used off-the-shelf components due to modest angular magnification and modest beam width provided to the back-end instruments by MAST. For DKIST, We have designed 2 custom achromats each to focus and to collimate the light. We are using off-the-shelf microlens array. We have also done the tolerance analysis over the radius of curvatures, tilt, and thickness and Abbe number. The future scope is to contact manufacturers with the calculated tolerances and further optimizations must be done as per manufacturing limitations. After the modifications instrument must be installed and calibrated.

# 3 Instrumentation

In the previous chapter, we have discussed the design aspects of pupil sampling Snapshot Spectrograph for MAST and DKIST. In this chapter, we discuss the realization of the snapshot spectrograph at MAST with the available components, the data reduction process and also about the snapshot spectro-polarimeter characterization at IIA Optics lab and its results.

## 3.1 Snapshot Spectrograph at MAST

We have used available components to set up the pupil sampling spectrograph at MAST and the following table lists the specifications. We have also characterized the Fabry Perot and the pre-filter.

Table 3.1 Optical Design specification of Snapshot Spectrograph installed at MAST

| Element | Specification |
| --- | --- |
| Telescope (T) Aperture | 90 mm |
| Telescope (T) Focal Length | 849.9 mm |
| Field Stop | 3 mm |
| L1 Focal Length | 150 mm |
| SH Lenslet Pitch | 1 mm |
| SH Lenslet Focal Length | 45 mm |
| L2 Focal Length | 180 mm |
| Fabry Perot Type | Air-Spaced Etalon |
| FP FSR and FWHM | 3.75 Å and 235 mÅ |
| L3 Focal Length | 300 mm |
| FOV | 50 arcsec |

## 3.2 Fabry Perot Etalon and Prefilter Calibration at MAST

To calculate the free spectral range and a full-width half maximum of FP and to find out the shift in the spectra, we had put the FP in front of the grating-based spectrograph and captured 11 frames with step size 100 from -500 to 500. Also, we have captured the pre-filter transmission profile which is plotted below.

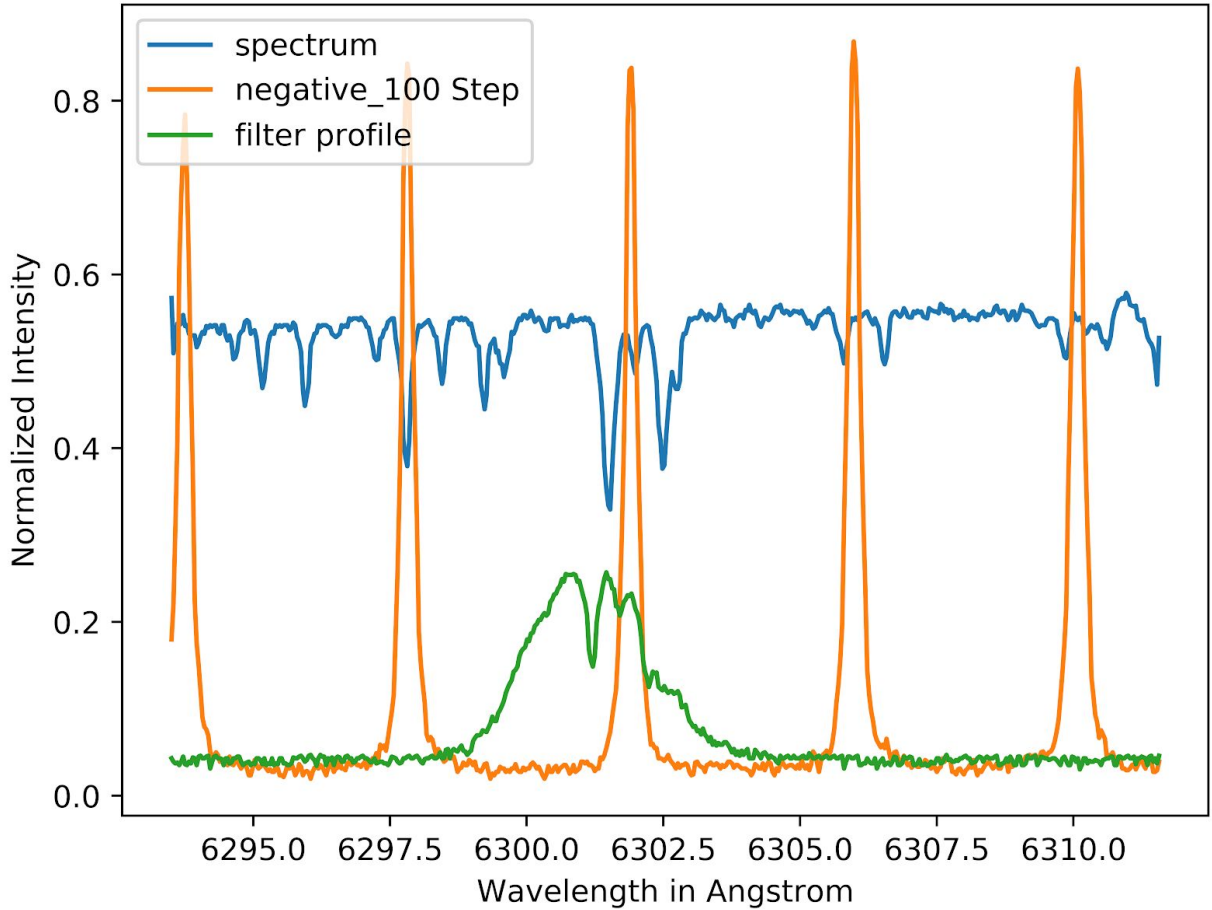

Fig 3.1 A sample FP profile at step -100 is overplotted with pre-filter profile and solar spectrum.

Table 3.2 FP Calibration Data

| Quantity | Values |
|---|---|
| Free Spectral Range | 3.75 Å |
| Full-Width Half Maximum | 235 mÅ |

Using the 11 data points about FP, we plotted the shift in spectra in wavelength we need to give to reference spectra (Step 0) for it to match the spectra at other steps. The graph shifts vs step number are plotted below.

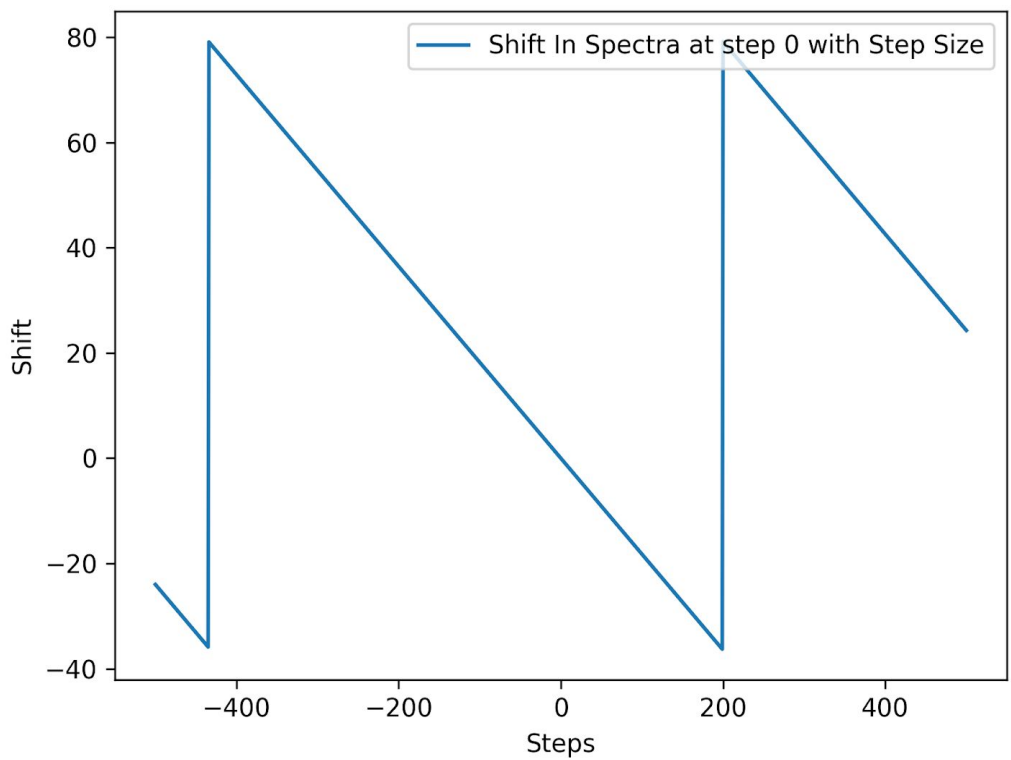

Fig 3.2 FP Shift in Spectra with respect to the spectrum at step 0 with step number

## 3.3 Instrument Setup

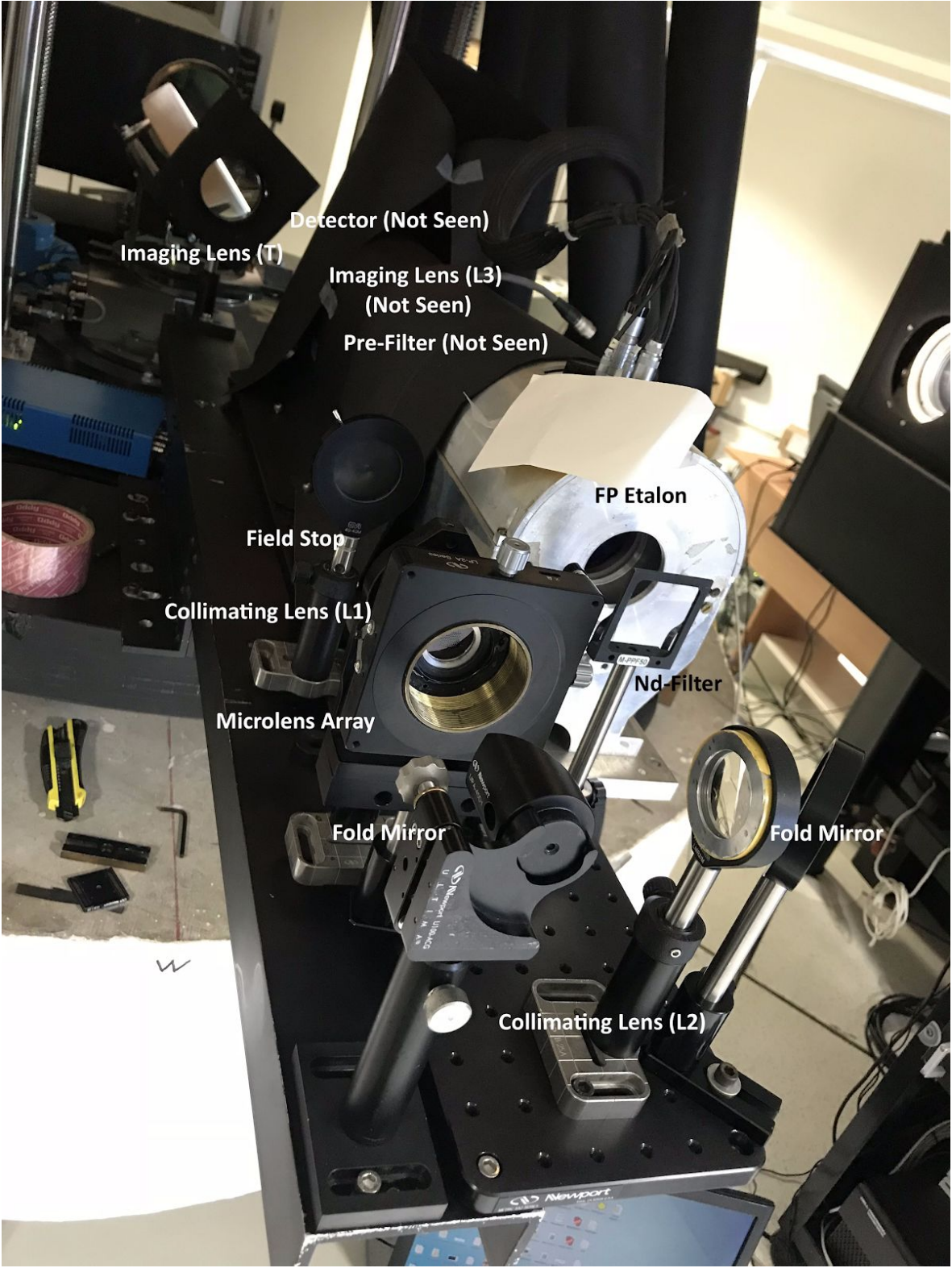

Fig 3.3 Photo Showing the experimental setup of Snapshot Spectrograph at MAST

Figure 3.3 shows the setup which consists of the imaging lens (T) followed by field stop. The beam is then collimated by the collimating lens (L1), which is making a pupil plane. The pupil plane is then sampled by microlens arrays. The fold mirror is folding the beam which is then collimated by collimating lens (L2). The beam is then again folded by fold mirror and passing through the Nd filter and then to the FP. the collimated beam from FP then passes through pre-filter followed by the imaging lens (L3) and gets imaged at the detector.

## 3.4 Data Reduction

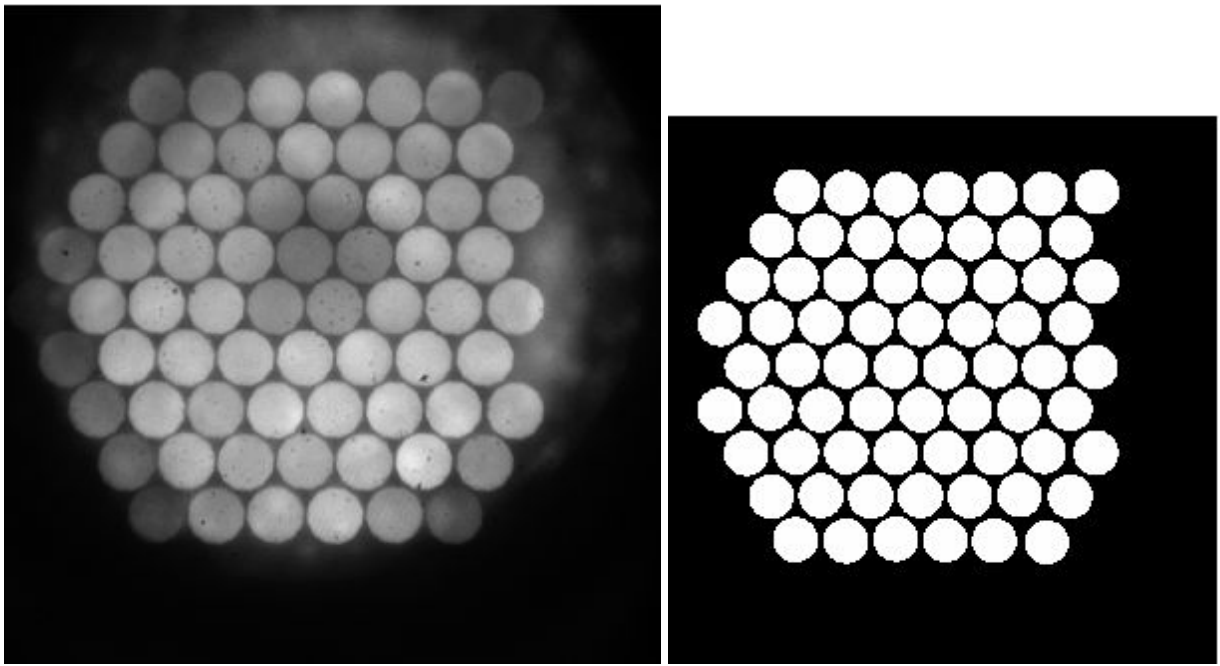

Fig 3.4 (a) spatio-spectral image plane (b) Mask for individual lenslet images by Hough Transform

A spatio-spectral image in x-y coordinates has a fixed wavelength for a given location $(x_i, y_i)$. Every $(x_i, y_i)$ corresponds to a spatial region on the ROI of $I_1$ (Figure 3.4 (a)). A data cube of $(x, y, \lambda)$ of the field can be built when spatially equivalent points $(x_n, y_n)$ are identified and arranged in increasing or decreasing order of wavelength. In other words, a spatial region (x, y) is present at each image of the lenslet in the spatio-spectral image plane, and each is at a different wavelength. To identify each lenslet image in the detector, Hough transform is used and a mask is generated. (Figure 3.4 (b))

## 3.4.1 Flat Fielding

Flat fielding a snapshot spatio-spectral image plane requires tuning the FP to change the transmitted wavelength. As the spacing of FP changes, the center wavelength shifts, and the continuum will shift from pixel to pixel. Sampling all the pixels at continuum and averaging will generate the master flat for the spatio-spectral plane.

Due to alignment errors in the setup, the detector's center might not coincide with the principal axis and the center wavelength($\lambda_0$) might be shifted from the spatio-spectral image plane center. To find the position of $\lambda_0$, we trace the path of absorption lines in flat scan data by choosing a row along the x-axis and a column along the y-axis. Stacking the rows or columns for all the images gives an arc. Choosing points on the arc and fitting a circle will give us the position of the center wavelength in the spatio-spectral image plane (Figure 3.5 (a) and (b))

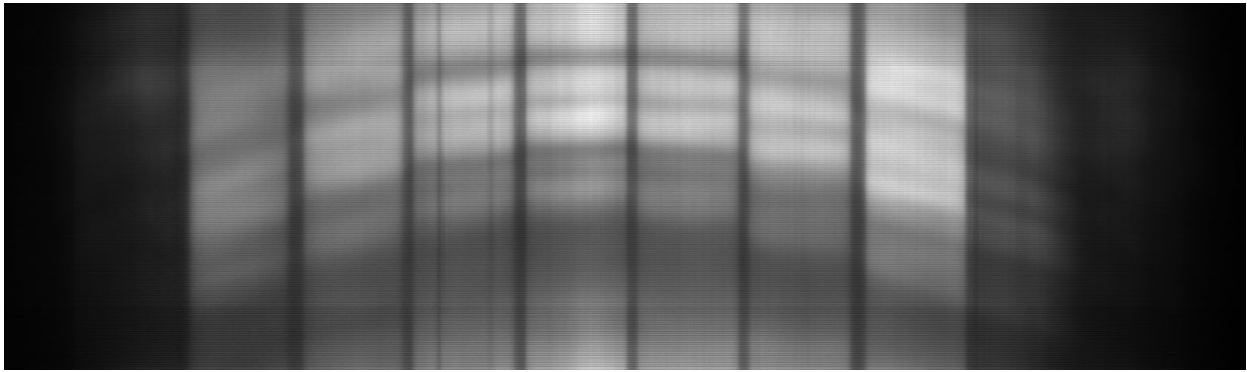
Fig 3.5 (a) Rows Stacked to trace the path of absorption lines

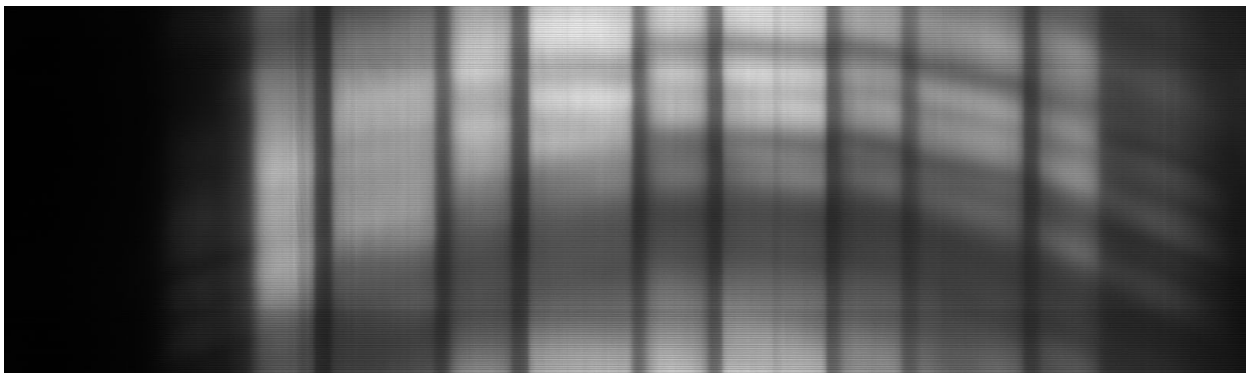
Fig 3.5 (b) Columns Stacked to trace the path of absorption lines

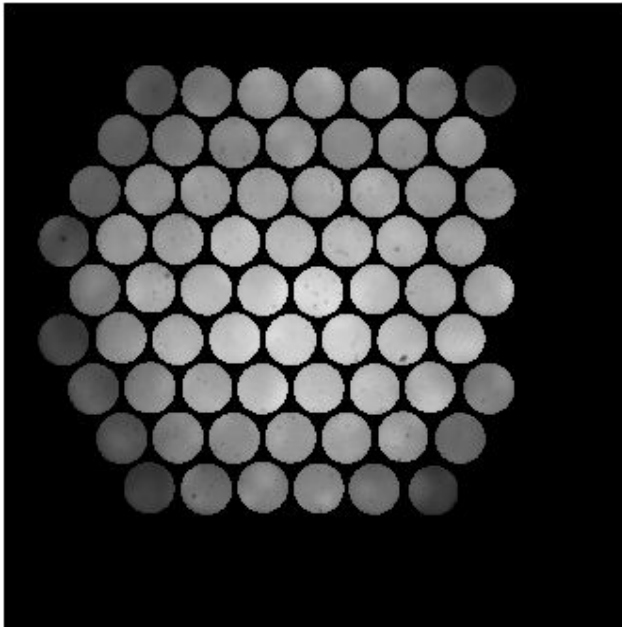
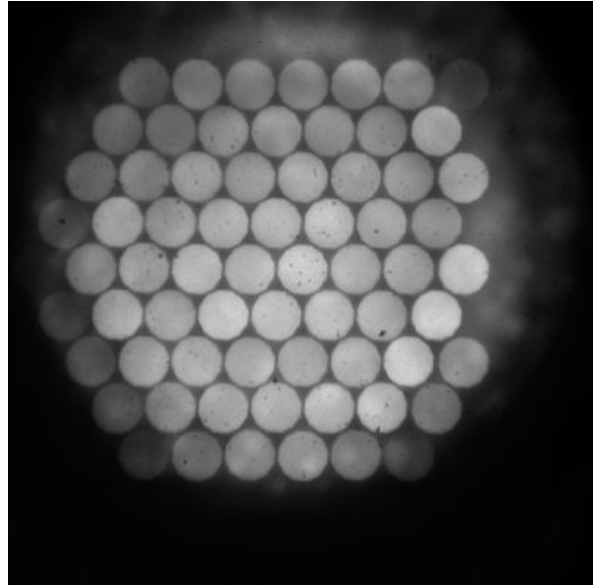

Fig 3.6 (a) Master Flat                           Fig 3.6 (b) Raw Data at step -150

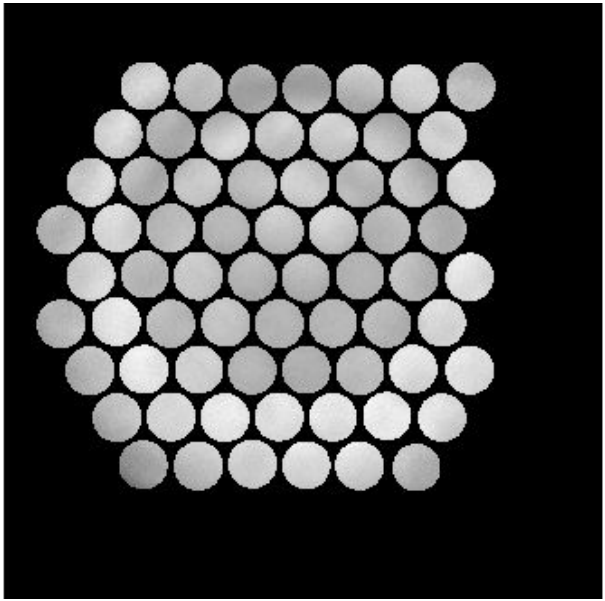

Fig 3.6 (c) Corrected Data at step -150

Figure 3.6 (a) shows the master flat generated from selecting pixels in the continuum wavelength region from the flat scans. Figure 3.6 (b) and (c) show a raw and a flat corrected image respectively.

## 3.4.2 Line Profile from Data

Due to bad observing conditions, we are unable to collect usable data. The data is with clouds passing by, which modulated the intensity and it is tough to separate cloud variations and variations due to FP.

We have plotted (Figure 3.7) the same graph of step -150 from the observed data. Although it does shows a reduced intensity at line 6301.5 Angstrom, it is not scientific data.

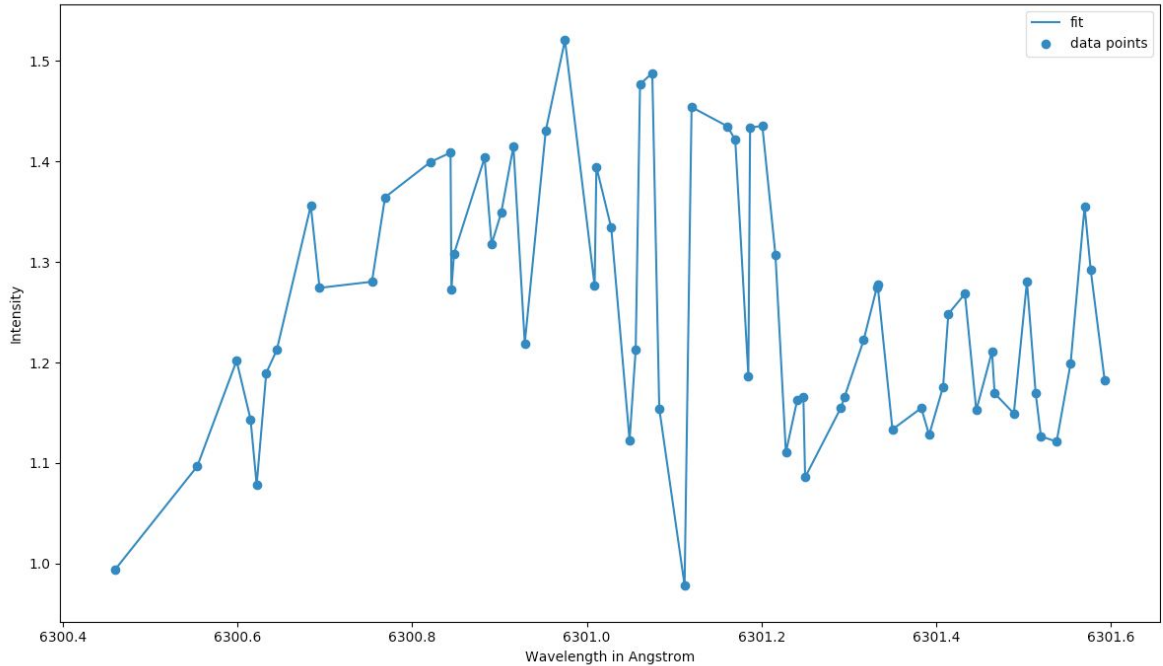

Fig 3.7 Line profile at a point using observed data at step -150

## 3.5 Simulated Data

We have calculated the wavelength incident at each pixel using the angle distribution at the detector (Figure 2.8) and equation 2.2. We have calculated the Intensity values for the wavelengths calculated using the BASS 2000 spectrum. Then we have multiplied the intensity values with the pre-filter profile and FP profile. At each step of the FP (spacing), the FP transmission is calculated using the shift vs step number plot (Figure 2.8). A sample simulated intensity profile is shown in Figure 3.8 for the FP step -150.

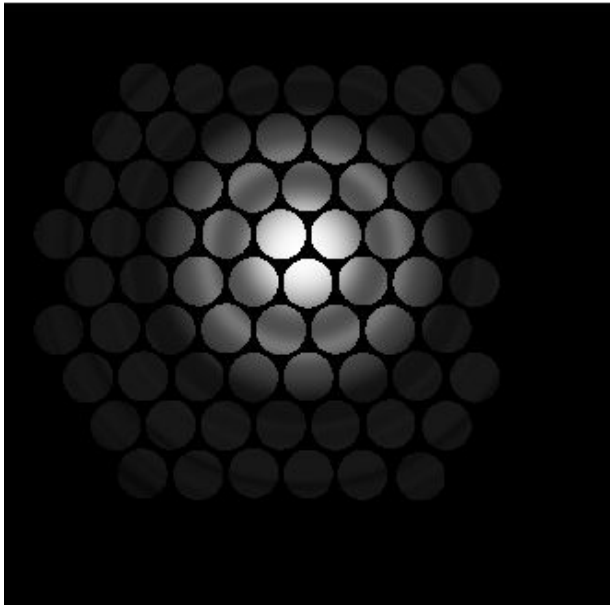

Fig 3.8 Simulated Data at step -150

## 3.5.1 Simulated Profile

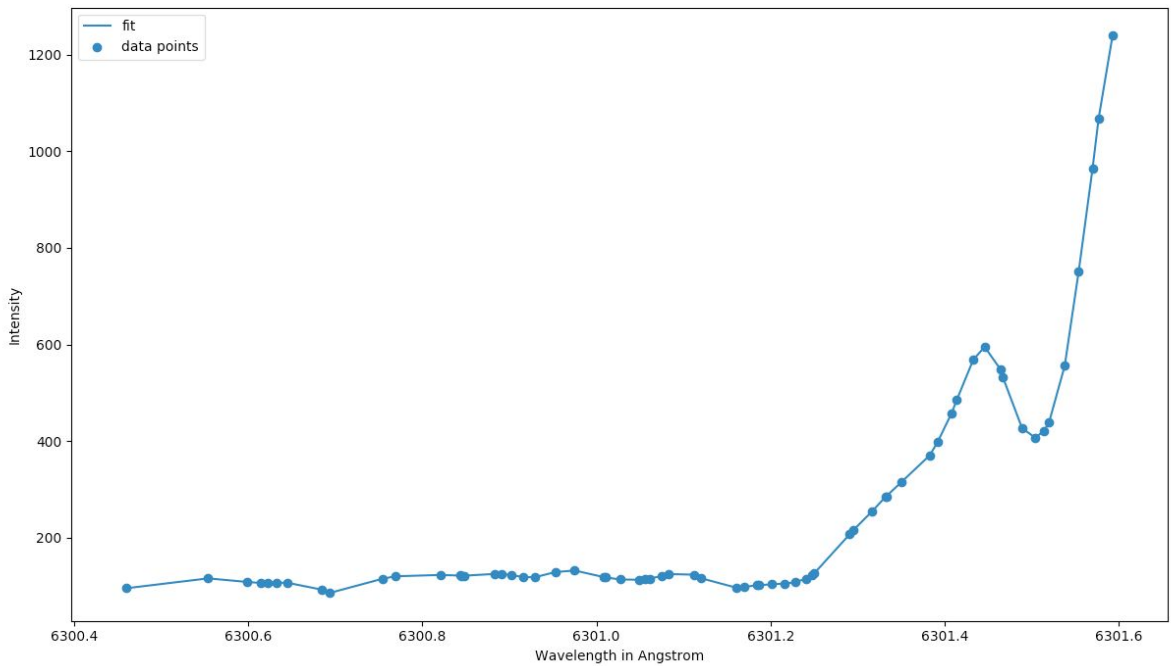

Fig 3.9 Line profile at a point using simulated data at step -150

Figure 3.9 displays a line profile at a spatial point using the simulated data. The filter response needs to be separated from this profile to get the appropriate line profile. We note that as expected line with center at 6301.538 is sampled with 8 wavelength points.

## 3.6 Snapshot Spectro-Polarimeter

We want to use this novel concept as a spectropolarimeter. Hence we had set up the snapshot spectro-polarimeter in IIA Optics lab and worked on the characterization of the polarimeter. We have used HeNe laser followed by a linear analyzer as a light source. The Linear analyzer is followed by a linear polarizer and a quarter wave plate being used together as Stokes polarimeter. It is then followed by a 50 μm pinhole. A 300 mm collimating lens then collimates the light and forms pupil plane at the focal plane. The microlens array is kept at the focal plane of the collimating lens which forms an image at 45 mm distance. The images from the microlens array are re-collimated by a 200 mm focal length lens, then the collimated beam is made to pass through the FP and reimaged by a 200 mm focal length lens at the detector. (Figure 3.10)

The following relations give the relation between intensity and the Stokes parameters.:

Intensity ($x^0$) signifies Intensity when the Linear polarizer of the polarimeter is kept $x^0$ from the horizontal.

Stokes I = Intensity($0^0$) + Intensity ($90^0$)
Stokes Q = Intensity ($0^0$) - Intensity ($90^0$)
Stokes U = Intensity ($45^0$) - Intensity ($135^0$)
Stokes V = Intensity (left circular) - Intensity (right circular)

### 3.6.1 Lab Snapshot Spectro-Polarimeter Setup

Table 3.3 Optical Design Specification of the Components used in IIA Optics Lab

| Element | Specification |
| --- | --- |
| Laser Type | HeNe (632.8 nm) |
| Pinhole Size | 50 μm |
| Collimating (L1) Lens Focal Length | 300 mm |
| SH Lenslet Pitch | 0.5 mm |
| SH Lenslet Focal Length | 45 mm |
| L2 Focal Length | 200 mm |
| Fabry Perot Type | Air-Spaced Etalon |
| FP FSR and FWHM | 5.64 Å and 1.41 Å |
| L3 Focal Length | 200 mm |

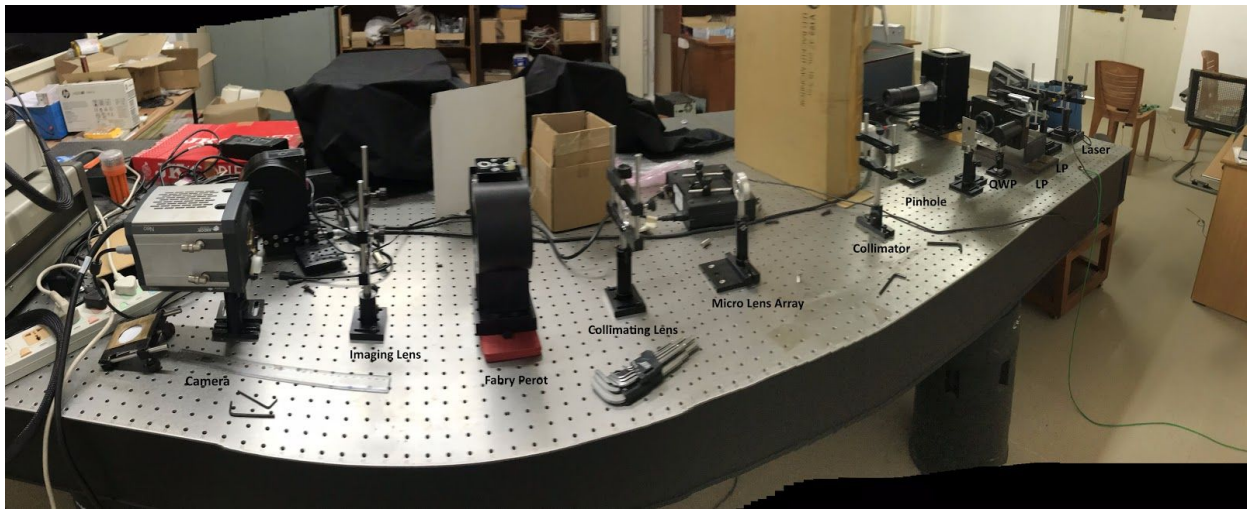

Fig 3.10 Snapshot Spectro-polarimeter Lab Setup

### 3.6.2 Data from Snapshot Polarimeter

We have collected data without FP and with FP to see the lenslet positions in the image plane.

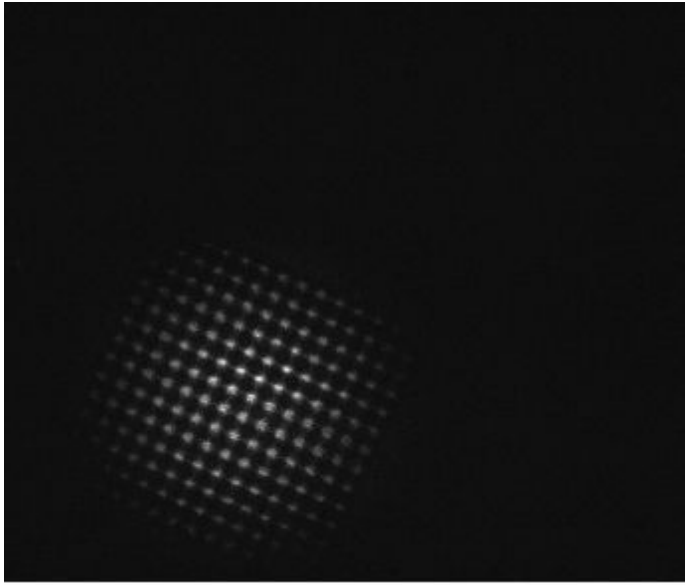

Fig 3.11 Lenslet images without FP

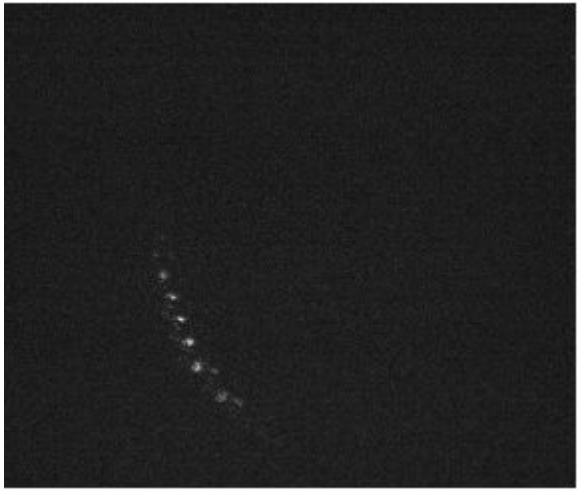

Fig 3.12 Data at Linear polarizer 0 degree orientation

Figure 3.11 shows the images of the lenslets when the FP is not present in the path. Hence we see all images from microlens array. Figure 3.12 shows the image from the snapshot spectrograph when the linear polarizer is at angle 0 degrees from the horizontal axis. We infer that the line is narrow and only the lenslets where HeNe wavelength(632.8 nm) wavelength fall upon are illuminated. In other words, The rest of the lenslet images are missing because they represent a wavelength not transmitted by the source. We also note that while taking these observations, we were unable to change X-parallelism of FP and hence unable to get it in operating mode. Hence the FP is used as is and FSR and FWHM can greatly differ than the values calculated earlier. For the same reason, we were unable to change the spacing too and hence cannot estimate the center wavelength with the step number and cannot produce line profiles as produced in the MAST data. Only polarimeter data is presented here.

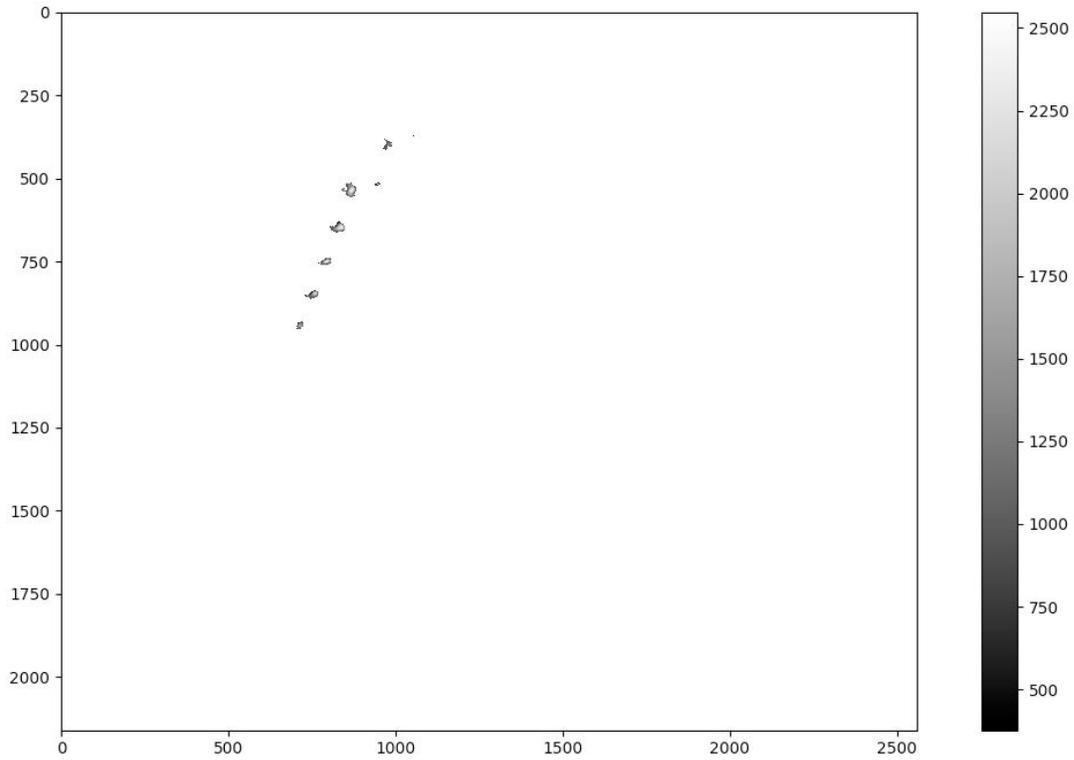

Fig 3.13 (a) Intensity Image at the detector for the linearly polarized He-Ne laser for the pixels at He-Ne passband.

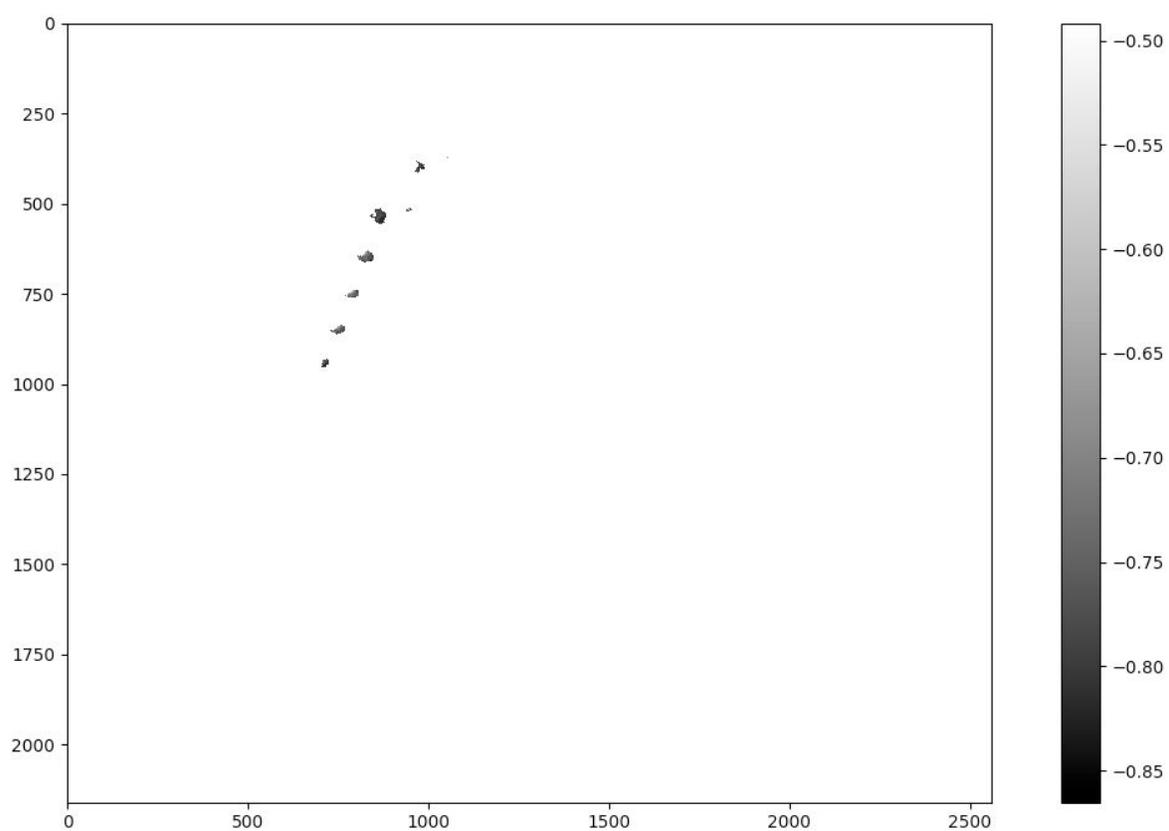

Fig 3.13 (b) Stokes Q/I for the linearly polarized He-Ne source at the detector for the pixels at He-Ne passband.

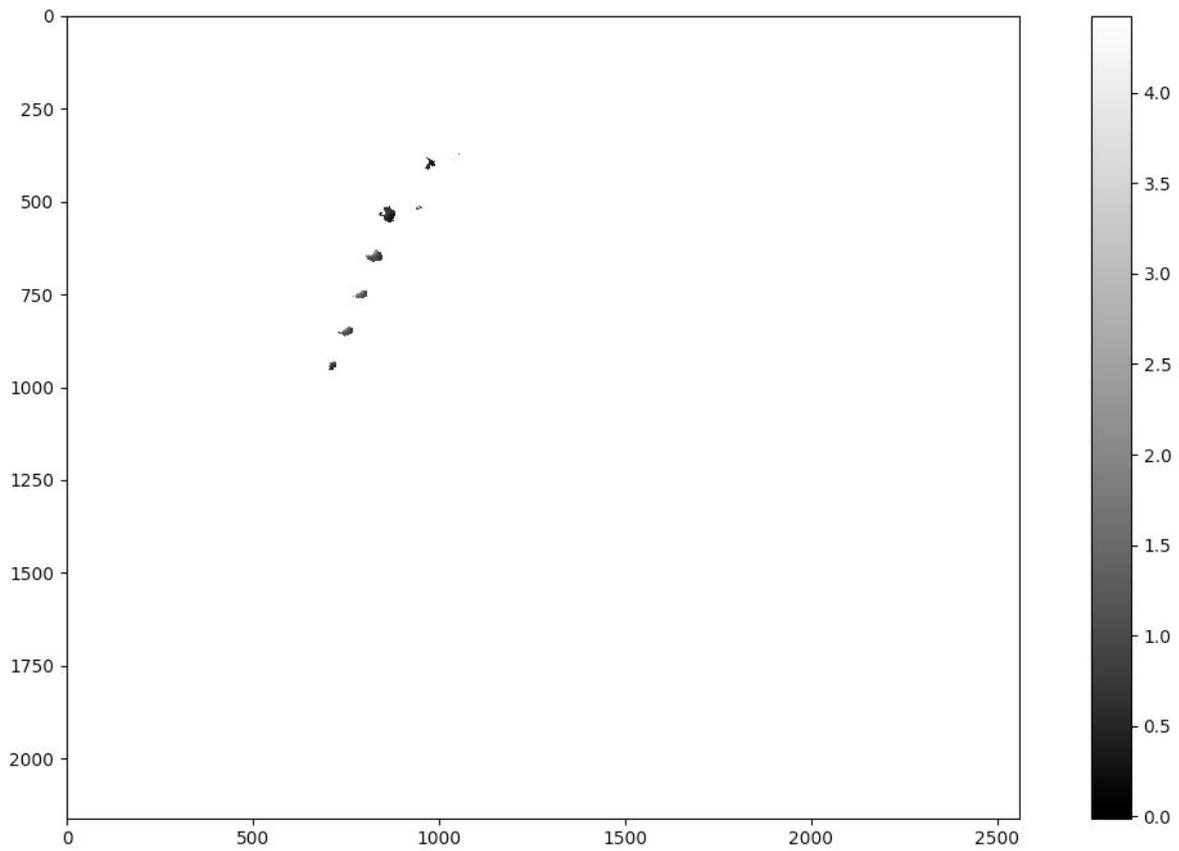

Fig 3.13 (c) Stokes U/I for the linearly polarized He-Ne source at the detector for the pixels at He-Ne passband.

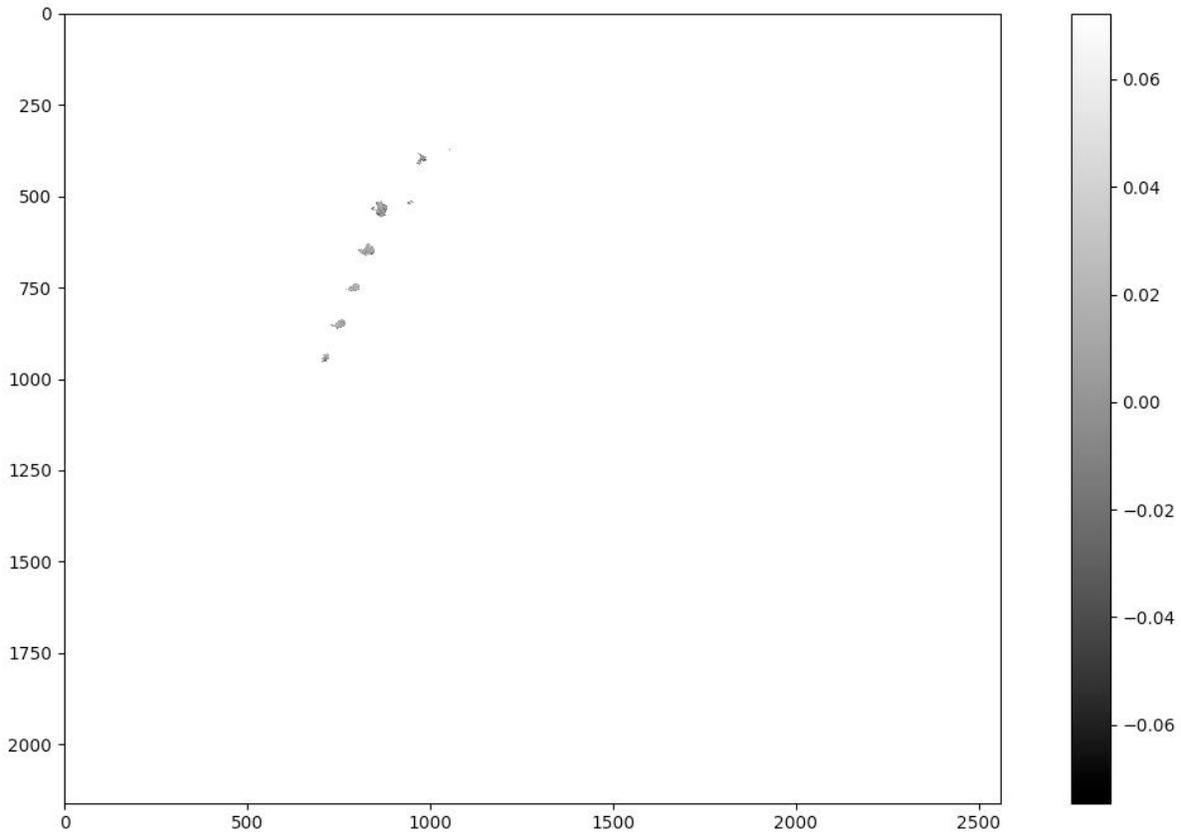

Fig 3.13 (d) Stokes V/I for the linearly polarized He-Ne source at the detector for the pixels at He-Ne passband.

The results (Figure 3.13 (a), (b), (c) and (d)) show Stokes Q as negative polarisation and Stokes U has a positive polarisation and Stokes V as zero. This indicates the light is coming at an angle given by $\tan^{-1}$(Stokes Q/Stokes U) to the horizontal axis and has no circular polarisation. The weak signal of Stokes V is present in the data due to inaccuracy in measurements which come while manually rotating the polarizer and waveplates.

## 3.7 Summary and Future Work

We have set up Snapshot Spectrograph at MAST Udaipur and Snapshot spectro-polarimeter at Optics Lab at IIA. The FP Calibration and observed profiles and simulated profiles are presented for the observed data. We have also simulated the BASS 2000 spectrum with prefilter and FP response and presented expected images and line profile. We have also set up the spectrograph and polarimeter at IIA Optics Lab and presented stokes parameter images. The initial results show no circular polarization and show that light is linearly polarized at an angle to the horizontal.

The future work is to perform the same experiment using sunlight as a source and rotating stages or LCVR based polarimeter and characterize it for it to be used with telescopes like MAST or DKIST.

# 4 Summary and Future Work

The snapshot spectroscopy is performed by sampling the pupil plane using the microlens array to get multiple images of the field of view. The multiple images of FOV are then collimated and made to pass through an FP and a pre-filter. Further, the beam is imaged using an imaging lens to get multiple images of FOV on the detector with a blue shift in the spectrum as we move radially outwards the optic axis.

Using the above concept, we have made optimized designs of pupil sampling Snapshot Spectrograph for Multi Application Solar Telescope (MAST) and Daniel K. Inouye Solar Telescope (DKIST). For the MAST we have used all off-the-shelf components which can be readily purchased. For the DKIST, we have designed 4 lenses (2 collimators, 2 imagers). We have also provided the optimum distance between various surfaces, and also have done tolerance analysis and presented the results. The future scope is to contact the manufacturers and further optimize the design with the inputs of manufacturing limitations. Both designs should be realized and calibrated.

We have also demonstrated the working of the snapshot spectrograph using the available components at MAST. We have calibrated the FP using a grating based spectrograph and calculated the free spectral range, full-width half maximum and the shift in reference spectra with the step number. We have done a full scan over the FP spacing and collected some data. We have generated the master flat and also calculated the line center at the detector plane. We have also used the BASS 2000 spectrum to simulate the expected data from the instrument and plotted both the observed and expected profiles.

We have also set up the instrument in the IIA optics lab for the purpose of understanding the polarimetry. We have used laser followed by a linear analyzer as a light source and used a linear polarizer followed by a quarter-wave plate as Stokes definition polarimeter. We have presented the initial results i.e. Stokes parameters from this experiment and demonstrated that there is no circular polarization. The future work is to perform the same experiment using sunlight as a source and rotating stages or LCVR based polarimeter and characterize it for it to be used with telescopes like MAST or DKIST.

# Appendix A

Analysis of Tolerances

File : E:\overall\before lenslet.zmx
Title: ATST_flat DM_f27
Date : 6/21/2019

Units are Millimeters.
All changes are computed using linear differences.

Paraxial Focus compensation only.

WARNING: Solves should be removed prior to tolerancing. Semi-diameters should be fixed.

WARNING: Boundary constraints on compensators will be ignored.

Criterion         : RMS Spot Radius in Millimeters
Mode              : Inverse Sensitivities, Limit Criterion Value : 0.01100000
Sampling          : 2
Nominal Criterion : 0.00302148
Test Wavelength   : 0.6328

Fields: XY Symmetric Angle in degrees

| # | X-Field | Y-Field | Weight | VDX | VDY | VCX | VCY |
|---|---|---|---|---|---|---|---|
| 1 | 0.000E+000 | 0.000E+000 | 4.000E+000 | 0.000 | 0.000 | 0.000 | 0.000 |
| 2 | 0.000E+000 | 4.125E-003 | 1.000E+000 | 0.000 | 0.000 | 0.000 | 0.000 |
| 3 | 0.000E+000 | -4.125E-003 | 1.000E+000 | 0.000 | 0.000 | 0.000 | 0.000 |
| 4 | 0.000E+000 | 5.893E-003 | 1.000E+000 | 0.000 | 0.000 | 0.000 | 0.000 |
| 5 | 0.000E+000 | -5.893E-003 | 1.000E+000 | 0.000 | 0.000 | 0.000 | 0.000 |
| 6 | 4.125E-003 | 0.000E+000 | 1.000E+000 | 0.000 | 0.000 | 0.000 | 0.000 |
| 7 | -4.125E-003 | 0.000E+000 | 1.000E+000 | 0.000 | 0.000 | 0.000 | 0.000 |
| 8 | 5.893E-003 | 0.000E+000 | 1.000E+000 | 0.000 | 0.000 | 0.000 | 0.000 |
| 9 | -5.893E-003 | 0.000E+000 | 1.000E+000 | 0.000 | 0.000 | 0.000 | 0.000 |

Sensitivity Analysis:

```
              |---------------- Minimum ---------------| |---------------- Maximum ---------------|
Type            Value       Criterion     Change          Value      Criterion      Change
TRAD  74      -0.12500000   0.00301723   -4.2534E-006    0.12500000  0.00302575    4.2722E-006
TRAD  75      -0.12500000   0.00303926    1.7785E-005    0.12500000  0.00300402   -1.7461E-005
TRAD  76      -0.12500000   0.00300402   -1.7454E-005    0.12500000  0.00303931    1.7828E-005
TFRN  77      -0.12500000   0.00300843   -1.3049E-005    0.12500000  0.00303473    1.3255E-005
TRAD  80      -0.12500000   0.00302795    6.4722E-006    0.12500000  0.00301506   -6.4189E-006
TRAD  81      -0.12500000   0.00303533    1.3856E-005    0.12500000  0.00300786   -1.3616E-005
TRAD  82      -0.12500000   0.00300747   -1.4006E-005    0.12500000  0.00303574    1.4257E-005
TRAD  88      -0.12500000   0.00303200    1.0519E-005    0.12500000  0.00301104   -1.0443E-005
TRAD  89      -0.12500000   0.00308440    6.2926E-005    0.12500000  0.00297244   -4.9042E-005
TRAD  90      -0.12500000   0.00298190   -3.9575E-005    0.12500000  0.00306223    4.0752E-005
TRAD  91      -0.12500000   0.00301456   -6.9152E-006    0.12500000  0.00303212    1.0645E-005
TRAD  92      -0.12500000   0.00303114    9.6585E-006    0.12500000  0.00301387   -7.6138E-006
TFRN  93      -0.12500000   0.00302249    1.0159E-006    0.12500000  0.00302047   -1.0119E-006
TTHI  74 75   -0.05000000   0.00301475   -6.7287E-006    0.05000000  0.00302827    6.7939E-006
TTHI  75 77   -0.05000000   0.00302805    6.5757E-006    0.05000000  0.00301497   -6.5113E-006
TTHI  76 77   -0.05000000   0.00302126   -2.1782E-007    0.05000000  0.00302170    2.1786E-007
TTHI  77 78   -0.05000000   0.00302148    3.3480E-016    0.05000000  0.00302148    1.7781E-016
```

| | | | | | |
|---|---|---|---|---|---|
| TTHI 78 79 | -0.05000000 | 0.00302148 | -5.8547E-017 | 0.05000000 | 0.00302148 -1.6610E-016 |
| TTHI 79 82 | -0.05000000 | 0.00302379 | 2.3127E-006 | 0.05000000 | 0.00301918 -2.2998E-006 |
| TTHI 80 82 | -0.05000000 | 0.00302356 | 2.0857E-006 | 0.05000000 | 0.00301940 -2.0769E-006 |
| TTHI 81 82 | -0.05000000 | 0.00302359 | 2.1073E-006 | 0.05000000 | 0.00301939 -2.0935E-006 |
| TTHI 82 83 | -0.05000000 | 0.00302148 | 1.2031E-012 | 0.05000000 | 0.00302148 -1.2039E-012 |
| TTHI 83 84 | -0.05000000 | 0.00301983 | -1.6474E-006 | 0.05000000 | 0.00302314 1.6610E-006 |
| TTHI 84 85 | -0.05000000 | 0.00302915 | 7.6727E-006 | 0.05000000 | 0.00301390 -7.5787E-006 |
| TTHI 85 86 | -0.05000000 | 0.00302148 | 9.5410E-018 | 0.05000000 | 0.00302148 -2.8189E-017 |
| TTHI 86 87 | -0.05000000 | 0.00302148 | 9.4542E-017 | 0.05000000 | 0.00302148 2.5934E-016 |
| TTHI 87 90 | -0.05000000 | 0.00301390 | -7.5775E-006 | 0.05000000 | 0.00302915 7.6706E-006 |
| TTHI 88 90 | -0.05000000 | 0.00302141 | -6.6431E-008 | 0.05000000 | 0.00302156 7.9691E-008 |
| TTHI 89 90 | -0.05000000 | 0.00301430 | -7.1756E-006 | 0.05000000 | 0.00302870 7.2241E-006 |
| TTHI 90 93 | -0.05000000 | 0.00302148 | -1.6985E-009 | 0.05000000 | 0.00302148 1.6987E-009 |
| TTHI 91 93 | -0.05000000 | 0.00301764 | -3.8405E-006 | 0.05000000 | 0.00302546 3.9776E-006 |
| TTHI 92 93 | -0.05000000 | 0.00302358 | 2.1041E-006 | 0.05000000 | 0.00301938 -2.0994E-006 |
| TEDX 74 75 | -0.05000000 | 0.00302620 | 4.7198E-006 | 0.05000000 | 0.00302624 4.7611E-006 |
| TEDY 74 75 | -0.05000000 | 0.00302434 | 2.8655E-006 | 0.05000000 | 0.00302331 1.8288E-006 |
| TETX 74 75 | -0.05000000 | 0.00303170 | 1.0221E-005 | 0.05000000 | 0.00301198 -9.4959E-006 |
| TETY 74 75 | -0.05000000 | 0.00302226 | 7.8140E-007 | 0.05000000 | 0.00302224 7.6312E-007 |

| | | | | | | |
|---|---|---|---|---|---|---|
| TEDX 76 77 | -0.05000000 | 0.00302506 | 3.5798E-006 | 0.05000000 | 0.00302503 | 3.5473E-006 |
| TEDY 76 77 | -0.05000000 | 0.00302315 | 1.6734E-006 | 0.05000000 | 0.00302372 | 2.2440E-006 |
| TETX 76 77 | -0.05000000 | 0.00301374 | -7.7425E-006 | 0.05000000 | 0.00303496 | 1.3485E-005 |
| TETY 76 77 | -0.05000000 | 0.00302428 | 2.8033E-006 | 0.05000000 | 0.00302433 | 2.8497E-006 |
| TEDX 80 82 | -0.01250000 | 0.00302167 | 1.9358E-007 | 0.02500000 | 0.00302229 | 8.1155E-007 |
| TEDY 80 82 | -0.05000000 | 0.00302135 | -1.2720E-007 | 0.05000000 | 0.00302183 | 3.4696E-007 |
| TETX 80 82 | -0.05000000 | 0.00302148 | 9.7709E-010 | 0.05000000 | 0.00302144 | -3.9326E-008 |
| TETY 80 82 | -0.05000000 | 0.00302156 | 8.4425E-008 | 0.05000000 | 0.00302157 | 9.3871E-008 |
| TEDX 88 90 | -0.05000000 | 0.00302562 | 4.1366E-006 | 0.05000000 | 0.00302566 | 4.1774E-006 |
| TEDY 88 90 | -0.05000000 | 0.00302940 | 7.9170E-006 | 0.05000000 | 0.00302185 | 3.6648E-007 |
| TETX 88 90 | -0.05000000 | 0.00306247 | 4.0995E-005 | 0.05000000 | 0.00304244 | 2.0961E-005 |
| TETY 88 90 | -0.05000000 | 0.00305262 | 3.1143E-005 | 0.05000000 | 0.00305251 | 3.1031E-005 |
| TEDX 91 93 | -0.05000000 | 0.00302147 | -1.0963E-008 | 0.05000000 | 0.00302147 | -1.0156E-008 |
| TEDY 91 93 | -0.05000000 | 0.00302146 | -1.9039E-008 | 0.05000000 | 0.00302148 | -2.7428E-009 |
| TETX 91 93 | -0.05000000 | 0.00306144 | 3.9960E-005 | 0.05000000 | 0.00304172 | 2.0238E-005 |
| TETY 91 93 | -0.05000000 | 0.00305173 | 3.0253E-005 | 0.05000000 | 0.00305163 | 3.0154E-005 |
| TSDX 74 | -0.05000000 | 0.00302198 | 5.0489E-007 | 0.05000000 | 0.00302200 | 5.1760E-007 |
| TSDY 74 | -0.05000000 | 0.00302167 | 1.9181E-007 | 0.05000000 | 0.00302179 | 3.1406E-007 |
| TIRX 74 | -0.00625000 | 0.00303111 | 9.6269E-006 | 0.00625000 | 0.00303116 | 9.6826E-006 |
| TIRY 74 | -0.02336744 | 0.00308974 | 6.8265E-005 | 0.02325541 | 0.00308771 | 6.6229E-005 |
| TSDX 75 | -0.05000000 | 0.00302360 | 2.1201E-006 | 0.05000000 | 0.00302363 | 2.1486E-006 |
| TSDY 75 | -0.05000000 | 0.00302311 | 1.6357E-006 | 0.05000000 | 0.00302196 | 4.7931E-007 |
| TIRX 75 | -0.00625000 | 0.00303139 | 9.9080E-006 | 0.00625000 | 0.00303133 | 9.8467E-006 |
| TIRY 75 | -0.02316361 | 0.00309449 | 7.3010E-005 | 0.02335861 | 0.00308588 | 6.4398E-005 |
| TSDX 76 | -0.05000000 | 0.00302506 | 3.5798E-006 | 0.05000000 | 0.00302503 | 3.5473E-006 |
| TSDY 76 | -0.05000000 | 0.00302315 | 1.6734E-006 | 0.05000000 | 0.00302372 | 2.2440E-006 |
| TIRX 76 | -0.00625000 | 0.00303810 | 1.6625E-005 | 0.00625000 | 0.00303817 | 1.6696E-005 |
| TIRY 76 | -0.02116722 | 0.00312363 | 0.00010215 | 0.02099356 | 0.00312666 | 0.00010519 |
| TSDX 77 | -0.05000000 | 0.00302148 | 0.00000000 | 0.05000000 | 0.00302148 | 0.00000000 |
| TSDY 77 | -0.05000000 | 0.00302148 | 0.00000000 | 0.05000000 | 0.00302148 | -1.3010E-018 |
| TIRX 77 | -0.00625000 | 0.00303778 | 1.6304E-005 | 0.00625000 | 0.00303772 | 1.6240E-005 |
| TIRY 77 | -0.02123760 | 0.00312044 | 9.8960E-005 | 0.02133479 | 0.00312712 | 0.00010564 |
| TSDX 80 | -0.02500000 | 0.00302159 | 1.1141E-007 | 0.02500000 | 0.00302160 | 1.2415E-007 |
| TSDY 80 | -0.05000000 | 0.00302160 | 1.2141E-007 | 0.05000000 | 0.00302146 | -2.0944E-008 |
| TIRX 80 | -0.00312500 | 0.00302171 | 2.3375E-007 | 0.00312500 | 0.00302173 | 2.5201E-007 |
| TIRY 80 | -0.05000000 | 0.00302745 | 5.9691E-006 | 0.05000000 | 0.00302906 | 7.5777E-006 |
| TSDX 81 | -0.02500000 | 0.00302136 | -1.1408E-007 | 0.02500000 | 0.00302136 | -1.2371E-007 |
| TSDY 81 | -0.05000000 | 0.00302171 | 2.3346E-007 | 0.05000000 | 0.00302335 | 1.8685E-006 |

| | | | | | |
|---|---|---|---|---|---|
| TIRX 81 | -0.00625000 | 0.00302132 | -1.6312E-007 | 0.00625000 | 0.00302133 -1.5201E-007 |
| TIRY 81 | -0.05000000 | 0.00304006 | 1.8580E-005 | 0.05000000 | 0.00304755 2.6069E-005 |
| TSDX 82 | -0.02500000 | 0.00302147 | -5.2807E-009 | 0.02500000 | 0.00302149 1.5498E-008 |
| TSDY 82 | -0.05000000 | 0.00302261 | 1.1296E-006 | 0.05000000 | 0.00302159 1.1075E-007 |
| TIRX 82 | -0.00156250 | 0.00302149 | 6.9753E-009 | 0.00156250 | 0.00302147 -4.7403E-009 |
| TIRY 82 | -0.05000000 | 0.00307616 | 5.4684E-005 | 0.05000000 | 0.00306715 4.5673E-005 |
| TSDX 88 | -0.05000000 | 0.00302491 | 3.4294E-006 | 0.05000000 | 0.00302491 3.4302E-006 |
| TSDY 88 | -0.05000000 | 0.00302279 | 1.3122E-006 | 0.05000000 | 0.00302702 5.5405E-006 |
| TIRX 88 | -0.05000000 | 0.00322331 | 0.00020183 | 0.05000000 | 0.00322332 0.00020184 |
| TIRY 88 | -0.05000000 | 0.00323856 | 0.00021708 | 0.05000000 | 0.00320768 0.00018620 |
| TSDX 89 | -0.05000000 | 0.00307499 | 5.3513E-005 | 0.05000000 | 0.00307500 5.3525E-005 |
| TSDY 89 | -0.05000000 | 0.00308429 | 6.2812E-005 | 0.05000000 | 0.00306563 4.4148E-005 |
| TIRX 89 | -0.05000000 | 0.00320825 | 0.00018677 | 0.05000000 | 0.00320823 0.00018675 |
| TIRY 89 | -0.05000000 | 0.00322499 | 0.00020351 | 0.05000000 | 0.00319123 0.00016975 |
| TSDX 90 | -0.05000000 | 0.00304763 | 2.6149E-005 | 0.05000000 | 0.00304766 2.6176E-005 |
| TSDY 90 | -0.05000000 | 0.00304406 | 2.2583E-005 | 0.05000000 | 0.00305120 2.9724E-005 |
| TIRX 90 | -0.05000000 | 0.00368812 | 0.00066664 | 0.05000000 | 0.00368800 0.00066652 |
| TIRY 90 | -0.05000000 | 0.00367204 | 0.00065057 | 0.05000000 | 0.00370369 0.00068221 |
| TSDX 91 | -0.05000000 | 0.00304664 | 2.5163E-005 | 0.05000000 | 0.00304663 2.5146E-005 |

| | | | | | | |
|---|---|---|---|---|---|---|
| TSDY 91 | -0.05000000 | 0.00303911 | 1.7635E-005 | 0.05000000 | 0.00305411 | 3.2632E-005 |
| TIRX 91 | -0.05000000 | 0.00322975 | 0.00020827 | 0.05000000 | 0.00322971 | 0.00020823 |
| TIRY 91 | -0.05000000 | 0.00325029 | 0.00022881 | 0.05000000 | 0.00320885 | 0.00018737 |
| TSDX 92 | -0.05000000 | 0.00304766 | 2.6176E-005 | 0.05000000 | 0.00304767 | 2.6193E-005 |
| TSDY 92 | -0.05000000 | 0.00305513 | 3.3649E-005 | 0.05000000 | 0.00304015 | 1.8675E-005 |
| TIRX 92 | -0.04000000 | 0.00322743 | 0.00020595 | 0.04000000 | 0.00322738 | 0.00020590 |
| TIRY 92 | -0.04000000 | 0.00324738 | 0.00022590 | 0.04000000 | 0.00320710 | 0.00018562 |
| TSDX 93 | -0.05000000 | 0.00302148 | 0.00000000 | 0.05000000 | 0.00302148 | 0.00000000 |
| TSDY 93 | -0.05000000 | 0.00302148 | -4.3368E-019 | 0.05000000 | 0.00302148 | 0.00000000 |
| TIRX 93 | -0.05000000 | 0.00380465 | 0.00078317 | 0.05000000 | 0.00380461 | 0.00078313 |
| TIRY 93 | -0.05000000 | 0.00377827 | 0.00075679 | 0.05000000 | 0.00383062 | 0.00080914 |
| TIRR 74 | -0.05000000 | 0.00306784 | 4.6358E-005 | 0.05000000 | 0.00297742 | -4.4058E-005 |
| TIRR 75 | -0.05000000 | 0.00297756 | -4.3921E-005 | 0.05000000 | 0.00306768 | 4.6206E-005 |
| TIRR 76 | -0.05000000 | 0.00307143 | 4.9955E-005 | 0.05000000 | 0.00297420 | -4.7276E-005 |
| TIRR 77 | -0.05000000 | 0.00297439 | -4.7084E-005 | 0.05000000 | 0.00307122 | 4.9741E-005 |
| TIRR 80 | -0.05000000 | 0.00302343 | 1.9544E-006 | 0.05000000 | 0.00301953 | -1.9479E-006 |
| TIRR 81 | -0.05000000 | 0.00302229 | 8.1070E-007 | 0.05000000 | 0.00302067 | -8.0970E-007 |
| TIRR 82 | -0.05000000 | 0.00301852 | -2.9563E-006 | 0.05000000 | 0.00302445 | 2.9705E-006 |
| TIRR 88 | -0.05000000 | 0.00304870 | 2.7225E-005 | 0.05000000 | 0.00299525 | -2.6227E-005 |

| | | | | | |
|---|---|---|---|---|---|
| TIRR 89 | -0.05000000 | 0.00302617 | 4.6890E-006 | 0.05000000 | 0.00301682 |
| -4.6579E-006 | | | | | |
| TIRR 90 | -0.05000000 | 0.00299080 | -3.0679E-005 | 0.05000000 | 0.00305353 |
| 3.2055E-005 | | | | | |
| TIRR 91 | -0.05000000 | 0.00304978 | 2.8304E-005 | 0.05000000 | 0.00299424 |
| -2.7235E-005 | | | | | |
| TIRR 92 | -0.05000000 | 0.00302968 | 8.2011E-006 | 0.05000000 | 0.00301337 |
| -8.1054E-006 | | | | | |
| TIRR 93 | -0.05000000 | 0.00298633 | -3.5149E-005 | 0.05000000 | 0.00305846 |
| 3.6979E-005 | | | | | |
| TIND 74 | -0.00025000 | 0.00309808 | 7.6599E-005 | 0.00025000 | 0.00295070 |
| -7.0779E-005 | | | | | |
| TIND 76 | -0.00025000 | 0.00297704 | -4.4435E-005 | 0.00025000 | 0.00306839 |
| 4.6915E-005 | | | | | |
| TIND 80 | -0.00025000 | 0.00300677 | -1.4707E-005 | 0.00025000 | 0.00303642 |
| 1.4939E-005 | | | | | |
| TIND 81 | -0.00025000 | 0.00302647 | 4.9931E-006 | 0.00025000 | 0.00301651 |
| -4.9650E-006 | | | | | |
| TIND 88 | -0.00025000 | 0.00296817 | -5.3308E-005 | 0.00025000 | 0.00307684 |
| 5.5357E-005 | | | | | |
| TIND 89 | -0.00025000 | 0.00304878 | 2.7299E-005 | 0.00025000 | 0.00299550 |
| -2.5984E-005 | | | | | |
| TIND 91 | -0.00025000 | 0.00302957 | 8.0898E-006 | 0.00025000 | 0.00301554 |
| -5.9364E-006 | | | | | |
| TIND 92 | -0.00025000 | 0.00301947 | -2.0051E-006 | 0.00025000 | 0.00302483 |
| 3.3537E-006 | | | | | |
| TABB 74 | -0.17099667 | 0.00307828 | 5.6804E-005 | 0.17099667 | 0.00296527 |
| -5.6212E-005 | | | | | |
| TABB 76 | -0.11989556 | 0.00296095 | -6.0532E-005 | 0.11989556 | 0.00308206 |
| 6.0585E-005 | | | | | |
| TABB 80 | -0.16554989 | 0.00302020 | -1.2828E-006 | 0.16554989 | 0.00302280 |
| 1.3214E-006 | | | | | |
| TABB 81 | -0.07133162 | 0.00302295 | 1.4668E-006 | 0.07133162 | 0.00302007 |
| -1.4044E-006 | | | | | |
| TABB 88 | -0.14497603 | 0.00321078 | 0.00018930 | 0.14497603 | 0.00286506 |
| -0.00015642 | | | | | |
| TABB 89 | -0.08062801 | 0.00285329 | -0.00016819 | 0.08062801 | 0.00322625 |
| 0.00020477 | | | | | |

```
TABB  91         -0.16041834    0.00318292    0.00016144    0.16041834    0.00287874
-0.00014274
TABB  92         -0.08062801    0.00286109   -0.00016038    0.08062801    0.00320395
0.00018247
```

Worst offenders:

| Type     | Value       | Criterion  | Change     |
|----------|-------------|------------|------------|
| TIRY  93 |  0.05000000 | 0.00383062 | 0.00080914 |
| TIRX  93 | -0.05000000 | 0.00380465 | 0.00078317 |
| TIRX  93 |  0.05000000 | 0.00380461 | 0.00078313 |
| TIRY  93 | -0.05000000 | 0.00377827 | 0.00075679 |
| TIRY  90 |  0.05000000 | 0.00370369 | 0.00068221 |
| TIRX  90 | -0.05000000 | 0.00368812 | 0.00066664 |
| TIRX  90 |  0.05000000 | 0.00368800 | 0.00066652 |
| TIRY  90 | -0.05000000 | 0.00367204 | 0.00065057 |
| TIRY  91 | -0.05000000 | 0.00325029 | 0.00022881 |
| TIRY  92 | -0.04000000 | 0.00324738 | 0.00022590 |

Estimated Performance Changes based upon Root-Sum-Square method:
Nominal RMS Spot Radius    :    0.00302148
Estimated change           :    0.00162596
Estimated RMS Spot Radius  :    0.00464744

Compensator Statistics:
Change in back focus:
Minimum            :    -0.421662
Maximum            :     0.421961
Mean               :     0.000007
Standard Deviation :     0.063596

Monte Carlo Analysis:
Number of trials: 20

Initial Statistics: Normal Distribution

| Trial | Criterion  | Change     |
|-------|------------|------------|
| 1     | 0.00408991 | 0.00106844 |
| 2 | 0.00408170 | 0.00106022 |
| 3 | 0.00386850 | 0.00084703 |
| 4 | 0.00344188 | 0.00042040 |
| 5 | 0.00463250 | 0.00161102 |
| 6 | 0.00380584 | 0.00078436 |
| 7 | 0.00370181 | 0.00068033 |
| 8 | 0.00436118 | 0.00133970 |
| 9 | 0.00372546 | 0.00070398 |
| 10 | 0.00380239 | 0.00078091 |
| 11 | 0.00413767 | 0.00111619 |
| 12 | 0.00494680 | 0.00192532 |
| 13 | 0.00335044 | 0.00032896 |
| 14 | 0.00423423 | 0.00121275 |
| 15 | 0.00351115 | 0.00048967 |
| 16 | 0.00511829 | 0.00209681 |
| 17 | 0.00394608 | 0.00092460 |
| 18 | 0.00351736 | 0.00049588 |
| 19 | 0.00395451 | 0.00093303 |
| 20 | 0.00413215 | 0.00111067 |

Number of traceable Monte Carlo files generated: 20

Nominal     0.00302148
Best        0.00335044   Trial   13
Worst       0.00511829   Trial   16
Mean        0.00401799
Std Dev     0.00046074

Compensator Statistics:
Change in back focus:
Minimum            :     -0.552143
Maximum            :      0.726911
Mean               :      0.010367
Standard Deviation :      0.317345

90% >    0.00478965
80% >    0.00429770
50% >    0.00395029
20% >    0.00360959

10% >     0.00347651

End of Run.